\documentclass[aps,amssymb,amsmath, twocolumn, pra, superscriptaddress,a4paper]{revtex4}
\usepackage{graphicx}
\usepackage{amsmath}    
\usepackage{bbm}    
\usepackage{amsfonts}
\usepackage[colorlinks, citecolor=red]{hyperref}
\usepackage{hyperref}
\usepackage{float}
\usepackage{color}
\usepackage{epsfig}
\usepackage{epstopdf}
\newcommand{\be}{\begin{equation}}
\newcommand{\ee}{\end{equation}}
\newcommand{\bc}{\begin{center}}
\newcommand{\ec}{\end{center}}
\newcommand{\bea}{\begin{eqnarray}}
\newcommand{\eea}{\end{eqnarray}}
\newcommand{\ba}{\begin{array}}
\newcommand{\ea}{\end{array}}

\usepackage{ulem}

\usepackage{amssymb}

\usepackage{mathrsfs}
\usepackage{xspace}
\usepackage{verbatim}
\usepackage{graphicx}
\usepackage{hyperref}
\usepackage{epstopdf}
\usepackage{verbatim}
\usepackage{braket}

\begin{document}

\title{Dynamics and energy spectra of aperiodic discrete-time quantum walks}
\author{C. V. Ambarish}
\affiliation{The Institute of Mathematical Sciences, C. I. T. Campus, Taramani, Chennai 600113, India}
\author{N. Lo Gullo}
\affiliation{Department of Physics, University of Milan, Italy}
\author{Th. Busch}
\affiliation{Okinawa Institute of Science and Technology Graduate University, 904-0495 Okinawa, Japan}
\author{L. Dell'Anna}
\affiliation{Department of Physics and Astronomy, University of Padua, Italy}
\author{C.M. Chandrashekar}
\affiliation{The Institute of Mathematical Sciences, C. I. T. Campus, Taramani, Chennai 600113, India}
\affiliation{Homi Bhabha National Institute, Training School Complex, Anushakti Nagar, Mumbai 400094,  India}

\begin{abstract}
Deterministically aperiodic sequences are an intermediary between periodic sequences and completely random sequences. Materials which are translationally periodic have Bloch-like extended states, while random media exhibit Anderson localisation. Materials constructed on the basis of deterministic aperiodic sequences such as Fibonacci, Thue-Morse, and Rudin-Shapiro exhibit different properties, which can be related to their spectrum. Here, by investigating the dynamics of discrete-time quantum walks using different aperiodic sequences of coin operations in position space and time we establish the role of the diffraction spectra in characterizing the spreading of the wavepacket.
\end{abstract}

\maketitle

%===================================
\section{Introduction}
%=====================================

\noindent Quantum walks are the quantum-mechanical analogue of classical random walks and just like their classical counterparts, the two variants exist: discrete-time quantum walks (DTQWs)~\cite{Aharonov1993} and continuous-time quantum walks (CTQWs)~\cite{Farhi1998}. They have played an important role in the field of quantum information processing as powerful parts of quantum algorithms~\cite{Kempe2003,Venegas-Andraca2012} and have been shown to be a universal quantum computation primitive; in that any quantum computation can be realized efficiently~\cite{Childs2009, Lovett2010}. Furthermore, with the ability to engineer quantum walks by controlling the parameters of the evolution operators, quantum effects such as localization~\cite{SCP11, Chandrashekar2011} and topological bound states~\cite{KRB10} have been simulated. Controlled dynamics of quantum walks have also allowed to simulate relativistic quantum dynamics where the speed of light is mimicked by the parameter in the evolution operator~\cite{Mey96, CBS10, MD13, MBD14, Cha13}. Their versatility and ease of implementation in physical systems such as, NMR ~\cite{Du2003,Ryan2005}, trapped ions ~\cite{Schimitz2009,Zahringer2010}, atoms ~\cite{Karski2009, Genske2013} and photonic systems~\cite{Do2005,Broome2010} makes them promising candidates to explore and engineer quantum dynamics beyond the Schr\"odinger equation.

Among the two variants of quantum walks, DTQWs are defined for a particle with an internal degree of freedom, to which the quantum coin operations are applied. These quantum coin operations can be position and time-dependent, which can lead to deviations from quadratic spreading of the wavepacket~\cite{NV01}, including the emergence of Anderson localisation~\cite{Anderson1958} for a completely disordered arrangement of coin operations in position space and/or  time~\cite{SCP11, Cha12}.  It is therefore natural to wonder about the behavior of DTQWs when the coin operation is aperiodic in nature~\cite{Ribeiro2004}. Studies of the behavior of DTQWs  for coin operations that are arranged in the aperiodic Fibonacci sequence~\cite{Ribeiro2004}  have already found a diffusive behavior without any signature of localization\cite{Romanelli2009,Ampadu2011}. However if this behavior is limited to just this sequence or extends to other aperiodic sequences as well is not yet clear. 

Below we will first briefly introduce the deterministic aperiodic sequences that we will study and that have previously been of great interest in condensed matter physics, photonics and material sciences~\cite{JoannopoulosReview1997, VardenyReview2008}.  

\begin{figure}[tb]
  \includegraphics[width=0.98\linewidth]{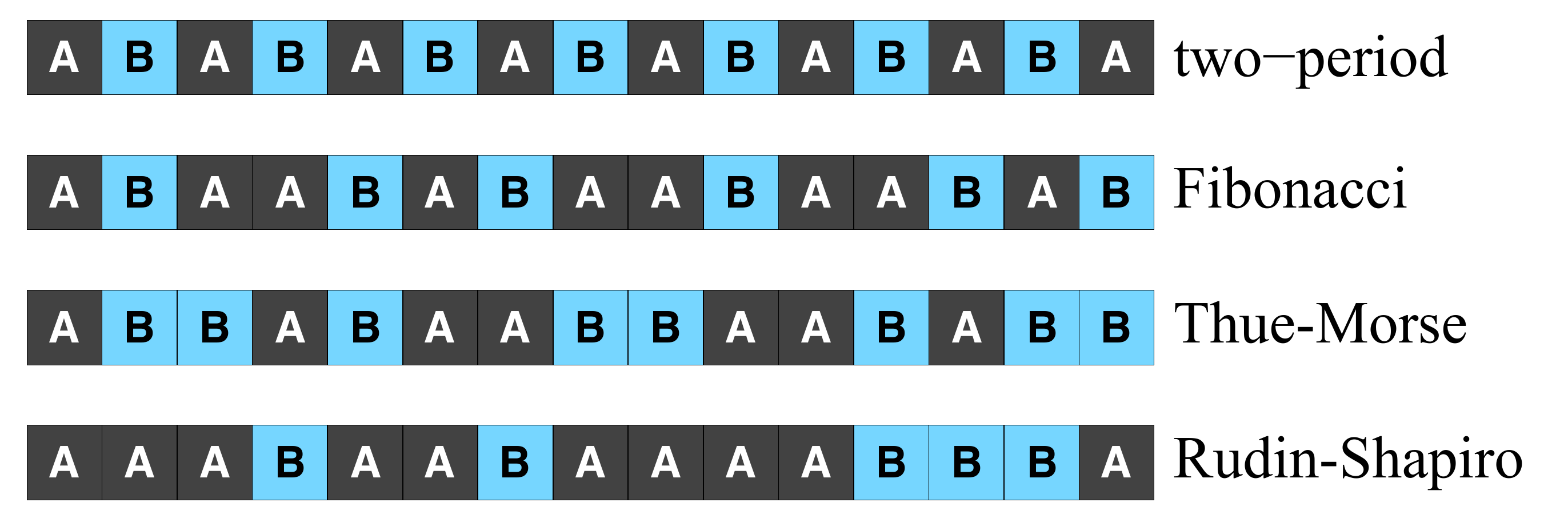}
  \caption{(Color online). Schematic of the arrangement of the two values for the two-periodic and the deterministic aperiodic sequences.  
  \label{fig:Schematic}}
\end{figure}

Materials with deterministic aperiodic sequences can be thought of as {\it in between} materials with long-range translational symmetry (crystals) and amorphous materials with no long range order.  Quasicrystals are one example~\cite{Shechtman1984} and searching for other non-periodic materials that have a crystal-like spectrum is an active research area. Of particular interest in this field have been structures based on the Fibonacci sequence~\cite{MerlinFibonacci1985, KohmotoFibonacci1983}, the Thue-Morse sequence~\cite{RiklundThueMorse1987} and the Rudin-Shapiro sequence~\cite{Vasconcelos1987}. The Fibonacci quasicrystal~\cite{MerlinFibonacci1985, KohmotoFibonacci1983}, a 1-D version of the celebrated Penrose tiles, has been studied in great detail because it has a pure-point spectrum. Thue-Morse crystals were shown to have a singular continuous spectrum~\cite{RiklundThueMorse1987}  and the Rudin-Shapiro crystals have a fully continuous spectrum~\cite{Vasconcelos1987}.  Several measures have been proposed to quantify the amount of order in aperiodic sequences, for instance based on entropy~\cite{Burrows1991,Berthe1994}, log-entropy~\cite{Levitov1988} and the structure of the energy-spectrum~\cite{Luck1989}. In  Fig.~\ref{fig:Schematic} we show a schematic of the different aperiodic sequences. 

Here we study the dynamics of the DTQW with spatial and temporal dislocations in the coin operator based on the three deterministically aperiodic sequences mentioned above: the quasiperiodic Fibonacci sequence, critical Thue-Morse sequence and the Rudin-Shapiro sequence. We compare our results with the ones for a standard two-periodic sequence of the form ABAB... and show that the support of the diffraction pattern plays a crucial role in determining the spreading properties of the quantum walker. By studying their energy spectra and their asymptotic behavior we show that the DTQW using the Fibonacci sequence, which has pure point spectrum, shows diffusive spreading behavior similar to the periodic walks. The Rudin-Shapiro sequence, which has a continuous spectrum, shows prominent localization behavior which is closer to a walk with a random coin sequence.  Quantum walks using the Thue-Morse sequence with a singular-continuous spectrum comprise of both, diffusing and localized component in the dynamics. 

\par The manuscript is organized as follows. In Section \ref{sec2} we briefly review the DTQW and in Section \ref{sec3} we introduce the different sequences of aperiodic walks with position and time dependent coin operations.   
In Section \ref{sec:en_spect}  and Section \ref{sec5} we present the study of energy spectra and the spreading of the walk for all configurations.  In Section \ref{sec:asymptotic} we present the asymptotic behaviour and conclude in Section \ref{conc}.

%=====================================
\section{Discrete-time Quantum Walk}
\label{sec2}
%=====================================
\noindent
The one-dimensional DTQW is defined on a position space and an internal (coin) space. 
The position space is spanned by the basis vectors $\ket{x}$, with $x \in \mathbb{Z}$, and the internal space of the particle is a two dimensional space spanned by $\ket{\uparrow} =  \left( \begin{array}{c} 1 \\ 0 \end{array} \right)$ and $\ket{\downarrow} =  \left( \begin{array}{c} 0 \\ 1 \end{array} \right)$. 
Each step of the walk consists of the application of one coin and one shift operator. 
The coin operator is unitary and can be written as
\begin{equation}
 \hat C(\theta)=\left( \begin{array}{ccc}
\cos(\theta)   &- \sin(\theta) \\
\sin(\theta)   & \cos(\theta) \end{array} \right).
\end{equation}
The shift operator performs the translation in the position space conditioned on
the internal degree of freedom of the walker and can be written as
\begin{equation}
 \hat S = \sum\limits_{x=-N}^N \ket{\uparrow}\bra{\uparrow} \otimes \ket{x-1}\bra{x} + \ket{\downarrow}\bra{\downarrow} \otimes \ket{x+1}\bra{x}.
\end{equation}
When the coin operation is homogeneous and time independent, the state of the system after $t$ steps ($t\in \mathbb{N}$) is given by 
$ \ket{\psi_{t}} = \hat W(\theta)\ket{\psi_{t-1}}$ where $ \hat W(\theta) =  \hat S( \hat I \otimes  \hat C(\theta))$.\\
In the following we will consider the situation where the coin operator can have two values, $\theta_1$ and $\theta_2$, and is either
time dependent but homogeneous in space or inhomogeneous in space and constant in time.  

%=====================================
\section{Aperiodic quantum walks with position and time dependent coin operations} 
\label{sec3}
%=====================================

In this section we will first introduce the sequences of the coin operations for the aperiodic settings we are going to study and then calculate the spread in the probability distribution after a certain time $t$. As a reference, we will also consider a DTQW with a regular distribution of two coins, which we will call a two-period sequence. For all walks we will fix $\theta_1 =\pi/4$ and study the dynamics for different values of $\theta_2$. 

The support of the diffraction pattern plays a crucial role in determining the spreading properties of the walker. According to the Lebesgue theory of measure the support can be decomposed  into a pure-point, singular continuous and absolutely continuous components. The Fibonacci, Thue-Morse and Rudin-Shapiro quasicrystals are particularly well suited to study the role of each of these components, since their diffraction measures
have pure-point, singular continuous and absolutely continuous support, respectively. On the other hand it has also been shown that their energy spectra are all singular continuous~\cite{Suto1989,Luck1989}.

%=====================================
\subsection{Periodic and aperiodic sequences}
%=====================================
{\it Two-periodic sequence:} 
\noindent The two-period sequence is of the form ABABAB........AB and it is the base sequence we will use to compare the deterministic aperiodic sequences to.  

% %=====================================
% \subsection{Fibonacci sequence}
% %=====================================
{\it Fibonacci sequence:} 
\noindent The Fibonacci sequence of two elements is defined by the substitution rule  
$A \rightarrow AB$ and $B \rightarrow A$, which can recursively be written as 
$S_{N+1} = S_NS_{N-1}$ $(S_1=B,S_2=A)$. 

% %=====================================
% \subsection{Thue-Morse sequence}
% %=====================================
{\it Thue-Morse:}
\noindent The Thue-Morse sequence of two elements is defined by the substitution rule 
$A \rightarrow AB$ and $B \rightarrow BA$ or recursively as $S_{N+1} = S_N\overline{S_N}$ $(S_1=A)$, where $\overline{S_N}$
is the string $S_N$ in which the two letters $A$ and $B$ have been exchanged.

% %=====================================
% \subsection{Rudin-Shapiro sequence}
% %=====================================
{\it Rudin-Shapiro:}
\noindent The Rudin-Shapiro sequence is defined by a four-element substitution sequence with the rules given by 
$P \rightarrow PQ $, $Q \rightarrow PR $, $R \rightarrow SQ $ and $S \rightarrow SR$. 
To obtain a sequence composed of only two elements, A and B, we perform the mapping
$(P,Q) \rightarrow A $, $(R,S) \rightarrow B$.

%====================================
\subsection{Position dependent coin operations}
%====================================
\noindent
By defining the states
$\ket{\nearrow_x}= \hat C(\theta(x))\ket{\uparrow}$ and $\ket{\swarrow_x}= \hat C(\theta(x))\ket{\downarrow}$,
we can rewrite the single step evolution operator $ \hat W= \hat S  \hat C'$ as
\begin{equation}
 \hat W=\sum\limits_{x=-N}^N \ket{\uparrow}\bra{\nearrow_x} \otimes \ket{x-1}\bra{x} + \ket{\downarrow}\bra{\swarrow_x} \otimes \ket{x+1}\bra{x}
\label{eq:WSingSp}.
\end{equation}
This allows us to interpret the evolution of the system
as an itinerant spin in an inhomogeneous magnetic field
and  we expect the mobility of the spin to be related to the 
spatial structure of the chosen sequence.
To see this let us consider the single step evolution 
operator $U=S\otimes C'$ where
\begin{eqnarray}
 \hat S &=& \sum\limits_{x=-N}^N \ket{\uparrow}\bra{\uparrow} \otimes \ket{x-1}\bra{x} + \ket{\downarrow}\bra{\downarrow} \otimes \ket{x+1}\bra{x}\nonumber\\ 
 \hat C' &=& \sum\limits_{x=-N}^N \ket{x}\bra{x} \otimes  \hat C(\theta(x))\nonumber.
\end{eqnarray}
Because $\theta(x)$ 
can only have two values, $\theta_1$ or $\theta_2$, we can partition the whole lattice 
$\mathcal{L}$ into two sublattices $\mathcal{L}_1$ and $\mathcal{L}_2$
such that $\mathcal{L}=\mathcal{L}_1\cup \mathcal{L}_2$ where $\mathcal{L}_i=\{x:\theta(x)=\theta_i\}$.

Using this observation we can rewrite
\begin{equation}
 \hat C' = \sum\limits_{x=-N}^N\ket{x}\bra{x}\otimes \left ( \hat {\overline{C}}+ w(x)\Delta  \hat C\right),
\end{equation}
where $\hat{\overline{C}}=(C(\theta_1)+ \hat C(\theta_2))/2$, 
$\Delta  \hat C=( \hat C(\theta_1)- \hat C(\theta_2))/2$ 
and $w(x)=1 \;\;\forall x\in \mathcal{L}_1$ and $w(x)=-1\;\; \forall x\in \mathcal{L}_2$.
The function $w(x)$ encodes the information on the geometry of the original spatial 
distribution of the two coins regardless of the actual form of the coin operators.

Specifically, in this form it is clear that the dynamics of the walker 
is linked to the underlying geometric distribution of the letters $A$ and $B$
of the sequence used. To see this let us introduce the states $\ket{k}$ such that $\braket{x|k}=e^{-\imath x k}/\sqrt{L}$ with $L=2N+1$ and the identity (for the spatial part) in this basis being $\hat {\bf 1}=\int dk \ket{k}\bra{k}$. This allows us to write
\begin{align}
 \hat C' =&\int dk\ket{k}\bra{k}\otimes  \hat{\overline{C}} \nonumber \\
              &+ \int dk dq\;\tilde f(k-q)\ket{k}\bra{q} \otimes \Delta C
\end{align}
where
\begin{eqnarray}
 \tilde f(q)&=&\frac{1}{L}\sum\limits_{x=-N}^N e^{\imath q x} w(x).
 \end{eqnarray}
The Fourier amplitudes $\tilde f(q)$ are nothing but the amplitudes of the diffraction
spectrum of the weighted Dirac comb $\omega=\sum\limits_x w(x)\delta_x$.

%=====================================
\subsection{Time dependent coin operations}
%=====================================
\noindent
To study the effect of temporally dependent coin operation we consider the application of $ \hat W(\theta(t))$
at the $t^\text{th}$ time step where $\theta(t)$ is chosen according to the distribution of $A\rightarrow \theta_1$
and $B\rightarrow \theta_2$ in the chosen sequence. For instance, if the chosen string is $ABABBBAABB$, then after $t=6$ time steps the evolution operator will be
$ \hat U= \hat W(\theta_2) \hat W(\theta_2) \hat W(\theta_2) \hat W(\theta_1) \hat W(\theta_2) \hat W(\theta_1)$.

Similar to the case of the position dependent coin we can define states
$\ket{\nearrow_t}= \hat C(\theta(t))\ket{\uparrow}$ and $\ket{\swarrow_t}= \hat C(\theta(t))\ket{\downarrow}$, which allows us to rewrite the single step evolution operator as
\begin{equation}
 \hat W(\theta(t)) = \sum\limits_{x=-N}^N \ket{\uparrow}\bra{\nearrow_t} \otimes \ket{x-1}\bra{x} + \ket{\downarrow}\bra{\swarrow_t} \otimes \ket{x+1}\bra{x} 
 \label{eq:WSingkick}
\end{equation}
The total evolution operator connecting the initial state
to the state at time step $t$, namely $ \hat U(t)=\prod\limits_{s=t}^{t} \hat W(\theta(s))$,
can be interpreted as the evolution of a spin in a time dependent magnetic field.
In this sense the change in coin operation for each time (change in magnetic field) in the system can be considered as a kicked itinerant spin on a lattice. 

%=====================================
\section{Energy spectra}
\label{sec:en_spect}
%=====================================

In this section we discuss the energy spectra  for the DTQWs using the different sequences introduced in preceding section. Here a comment is necessary on what we mean by energy  spectrum in this context and how we derive it. In general an operator of the form given in Eq.~\eqref{eq:WSingSp} acting on a finite space is not a unitary operator due to the spatial part. The iterative application of the single step operators therefore results, 
in general, in a non-probability conserving evolution under a total operator $\hat T$. The latter is such that $\hat T \hat T^{\dag}=\hat T^{\dag}\hat T=\hat I$ still acts as a unitary operator in a sub lattice $[-\tilde N,\tilde N]$
with $\tilde N<N$. Therefore, for localized initial states and for a certain $t^*$ such that the probability $p_x=|\braket{x,\uparrow|\psi(t^*)}|^2+|\braket{x,\downarrow|\psi(t^*)}|^2$ of finding the walker in at a given position $x$ is $p_x=0$ if $x\notin[-\tilde N,\tilde N]$,  we can restrict the dynamics of the walker to that interval and consider it as a unitary evolution. The energy spectrum is then the spectrum of the unitary operator $\hat U$
which is the restriction of $\hat T$ acting onto the above mentioned subspace. It is worth stressing that this is mainly due to the numerical limitations, whereas this problem does not arise in the thermodynamic limit.

\begin{figure}[tb]
 \includegraphics[width=8cm]{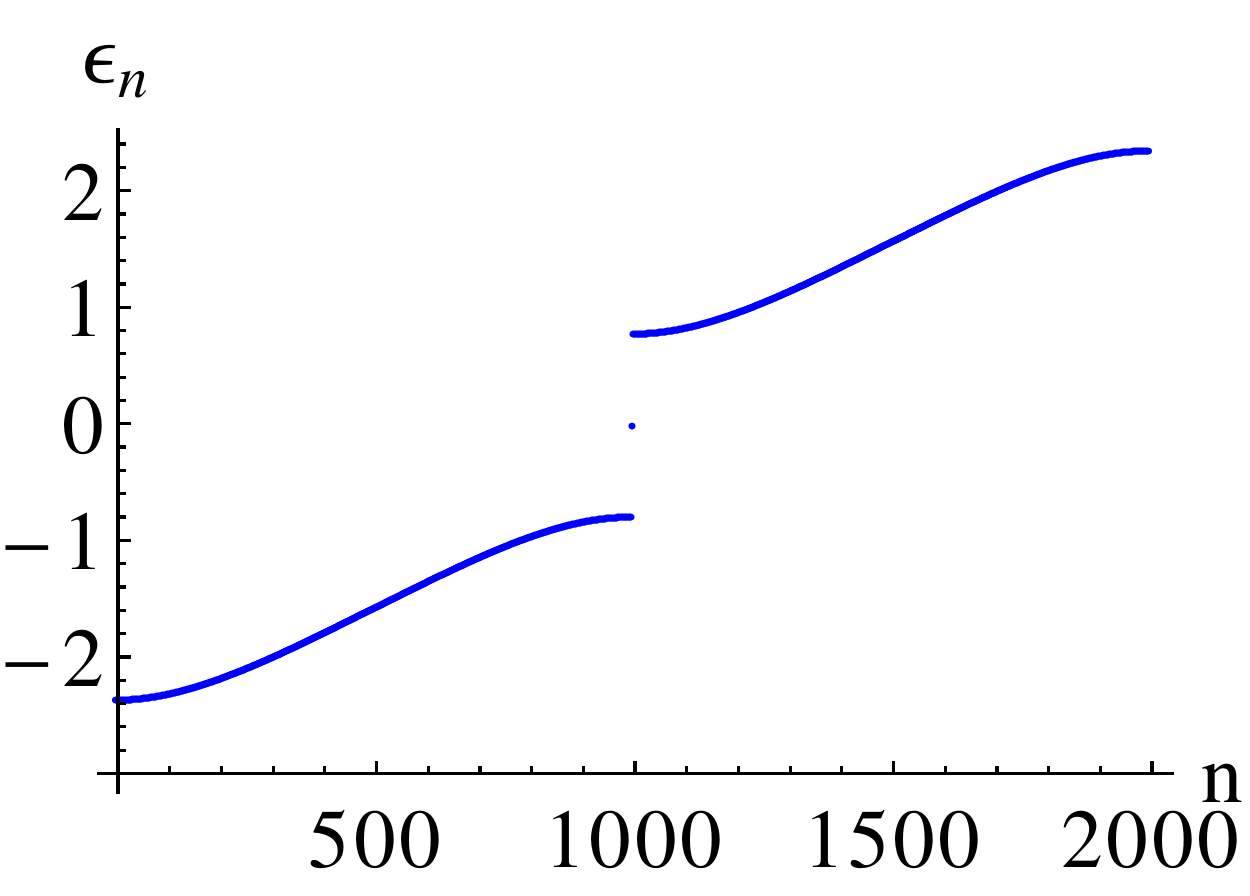}
\caption{(Color online). Spectrum $\epsilon_n=\imath \log(\lambda_n)$
where $\lambda_n$  are the eigenvalues of the operator $ \hat W$ in Eq.~\eqref{eq:WSingSp}
for the case $\theta_1=\theta_2=\pi/4$.}
\label{fig:hom_spect}
\end{figure}

The energy spectrum of a homogeneous and time-independent DTQW for the case $\theta=\pi/4$ is shown in Fig.\,\ref{fig:hom_spect}. It is given by $\epsilon_n=\imath \log(\lambda_n)$, with the $\lambda_n$ being the eigenvalues of $\hat W(\theta)$, and can be seen to consist of two main bands, each of which is the spectrum of a tight binding Hamiltonian. 

\begin{figure}[tb]
 \includegraphics[width=8cm]{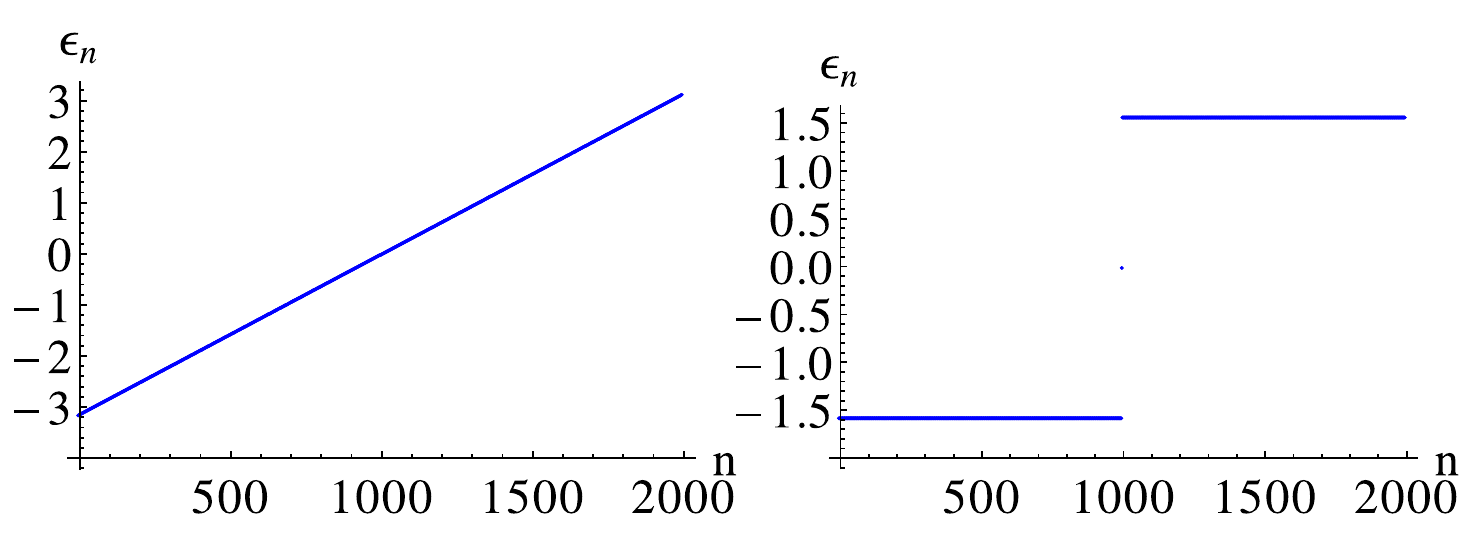}
\caption{(Color online). Energy spectra $\epsilon_n=\imath \log(\lambda_n)$ where $\lambda_n$ are the eigenvalues of the operator $ \hat W(\theta)$ in Eq.~\eqref{eq:WSingSp}, for $\theta=0$ (left) and $\theta=\pi/2$ (right).}
\label{fig:hom_spect_0_pi2}
\end{figure}

The two bands are due to the presence of the internal degree of freedom (coin) and the two extreme cases are ${\bf (a)}$ $\theta=n \pi$ and ${\bf (b)}$ $\theta=(2n-1) \pi/2$ which correspond to $C=(-1)^n \hat I$ and $C=(-1)^n\left(\ket{\uparrow}\bra{\downarrow}-\ket{\downarrow}\bra{\uparrow}\right)$,
respectively. To see this let us consider an infinite lattice $(N\rightarrow \infty)$ and introduce the states 
$\ket{k}$ such that $\braket{x|k}=e^{-\imath x k}/\sqrt{L}$ with $L=2N+1$ and the identity (for the spatial part) 
in this basis $\hat {\bf 1}=\int dk \ket{k}\bra{k}$. This allows us to rewrite the single step evolution operator as
\begin{equation}
\hat W= \int dk\ket{k}\bra{k}\otimes (e^{-\imath k}\ket{\uparrow}\bra{\uparrow}+e^{\imath k}\ket{\downarrow}\bra{\downarrow}) \hat C(\theta).
\end{equation}
For the case ${\bf (a)}$ $C(n\pi)=\pm\hat{I}_C$  the eigenstates of $\hat W$ are $\ket{k}\ket{\uparrow}$ and $\ket{k}\ket{\downarrow}$ with eigenvalues $e^{-\imath k}$ and $e^{\imath k}$ respectively.
On the other hand in case ${\bf (b)}$ $\hat C=(-1)^n\left(\ket{\uparrow}\bra{\downarrow}-\ket{\downarrow}\bra{\uparrow}\right)$
the eigenstates are given by $e^{\pm\imath\frac{\pi}{2}}\int dk\;\ket{k}(e^{\imath \frac{k}{2}}\ket{\uparrow}\pm e^{-\imath \frac{k}{2}}\ket{\downarrow})/\sqrt{2}$ with eigenvalues $\pi/2$ and $-\pi/2$.
Therefore the spectrum in case ${\bf (a)}$  is linear without gap at $\epsilon=0$ 
whereas in case ${\bf (b)}$ it is flat inside the two bands which are separated by a gap $\Delta=\pi$
as shown in Fig.\,\ref{fig:hom_spect_0_pi2}.

When it comes to the study of the energy spectra of a DTQWs the distinction between spatial aperiodic sequences and temporal aperiodic ones is important.

%=====================================
\subsection{Position dependent coin operations}
%=====================================

\begin{figure}[tb]
\includegraphics[width=8cm]{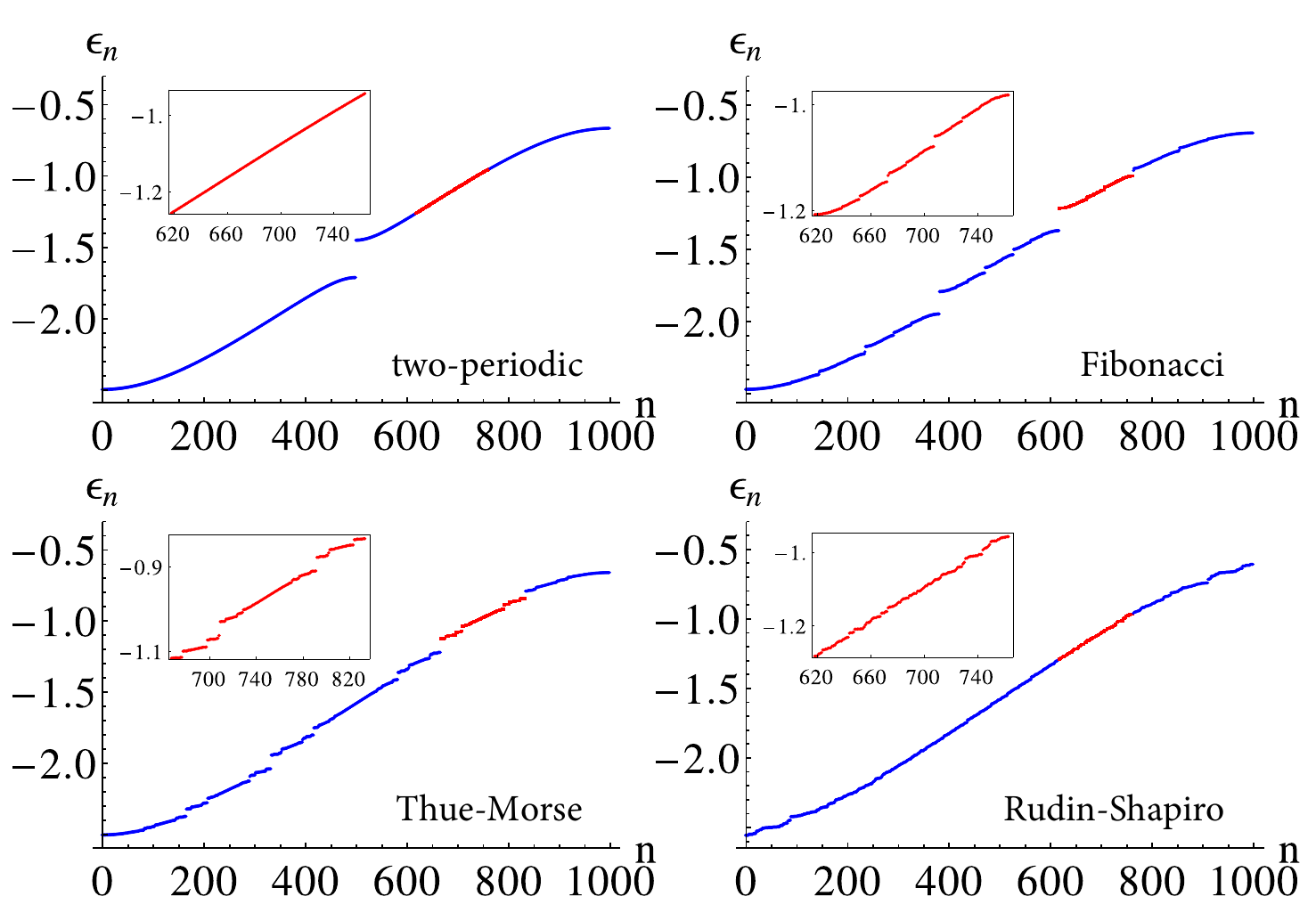}
\caption{(Color online). Lowest band of the spectrum $\epsilon_n=\imath \log(\lambda_n)$ where the $\lambda_n$ are the eigenvalues of the operator $ \hat W(\theta)$ in Eq.~\eqref{eq:WSingSp}. The angles chosen are $\theta_1=\pi/4$ and $\theta_2=\pi/6$ and are distributed (from top left to bottom right) according to the two-periodic, the Fibonacci, the Thue-Morse and the Rudin-Shapiro sequence. In the inset of each figure we show a zoomed in part of the spectrum to show the self-similarity of the spectra and the presence of gaps at all energies.}
\label{fig:inhom_spect}
\end{figure}

In the case of DTQWs with position dependent coin operations $(\theta_2\neq \theta_1)$ 
distributed according to a given aperiodic sequence the spectrum of the system is the same at all time steps 
and coincides with that of the single step evolution operator $\hat W(\theta)$.
It can be seen from Fig.~\ref{fig:inhom_spect} that the effect of different coin operations on different sites is to open new gaps inside the two above mentioned main energy bands thus creating sub-bands.
The position of these new gaps can be determined in first order perturbation theory
(i.e. considering the case $|\theta_2-\theta_1|\approx 0$) by finding that they 
open where the intensity of the Fourier transform of $w(x)$ is higher \cite{Suto1989,Luck1989,Vignolo2014,Nico2016}.  Within the two main bands the  structure is 
therefore determined by the spatial distribution of the two coins and consequently
by the geometry of the chosen arrangement. 

\begin{figure}[tb]
\includegraphics[width=8cm]{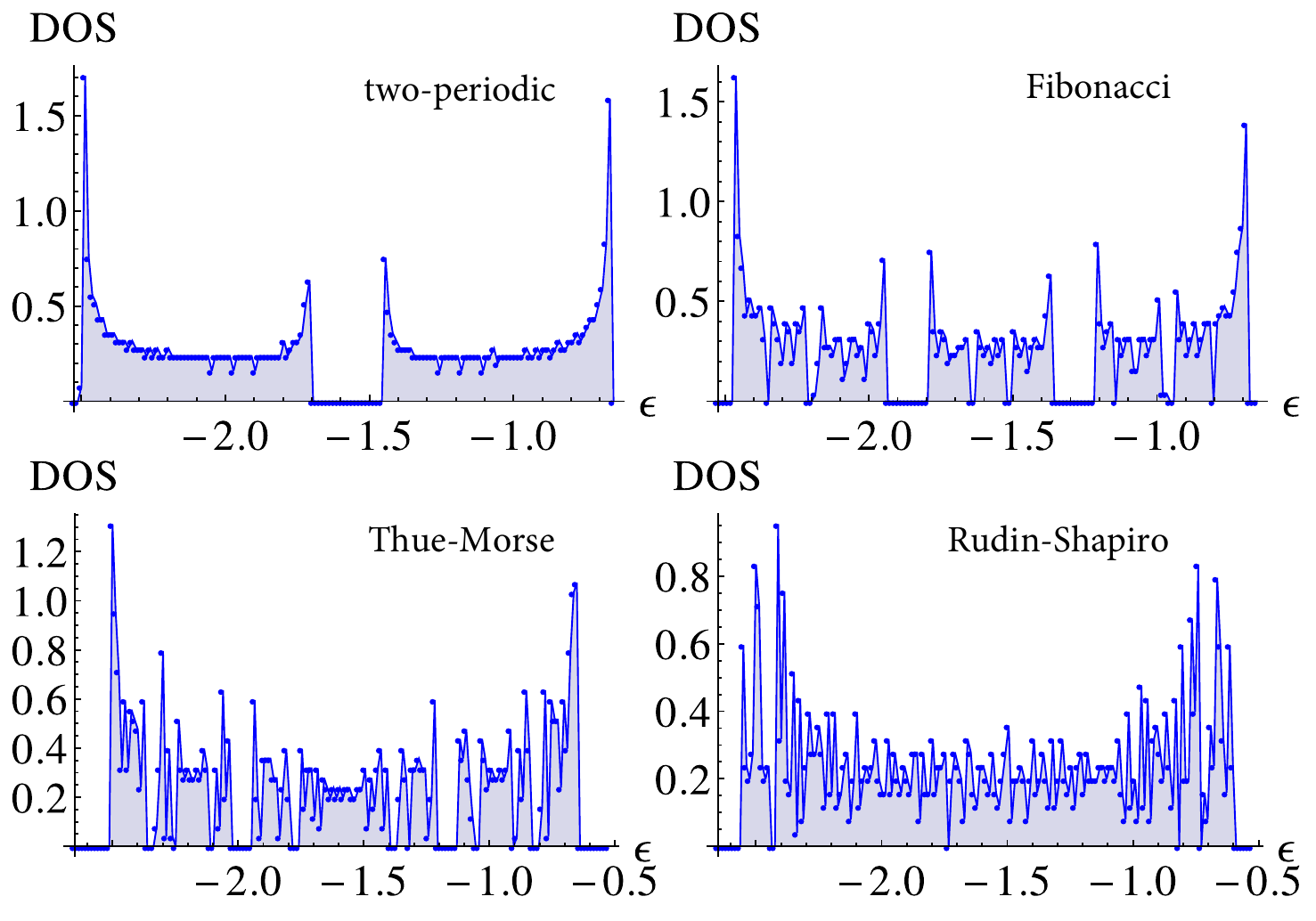}
\caption{(Color online). Density of states normalized to the total number of states 
corresponding to the spectra in Fig.~\ref{fig:inhom_spect}. The energy window used to compute these graphs
is the same as the one that will be used for the analysis of the asymptotic properties $d\epsilon=2\pi/t_f$
where $t_f=500$ is the number of steps.}
\label{fig:dos_space}
\end{figure}

The above spectra give rise to the densities of states (DOS) shown in Fig.~\ref{fig:dos_space},
from which key properties of each distribution can be identified.
In particular we notice the self-similarity of the energy spectrum of the Fibonacci sequence
(top-right) and the broad availability of states for the Rudin-Shapiro sequence throughout
the whole sub-band. We also notice that each DOS for all aperiodic arrangements shows singular
features (spikes in the DOS), which are a result of the opening
of the gaps in the spectrum on a set of zero measure \cite{Suto1989,Luck1989}.
This translates into plateaus in the integrated DOS.

%=====================================
\subsection{Time dependent coin operations}
%=====================================

\begin{figure}[tb]
 \includegraphics[width=8cm]{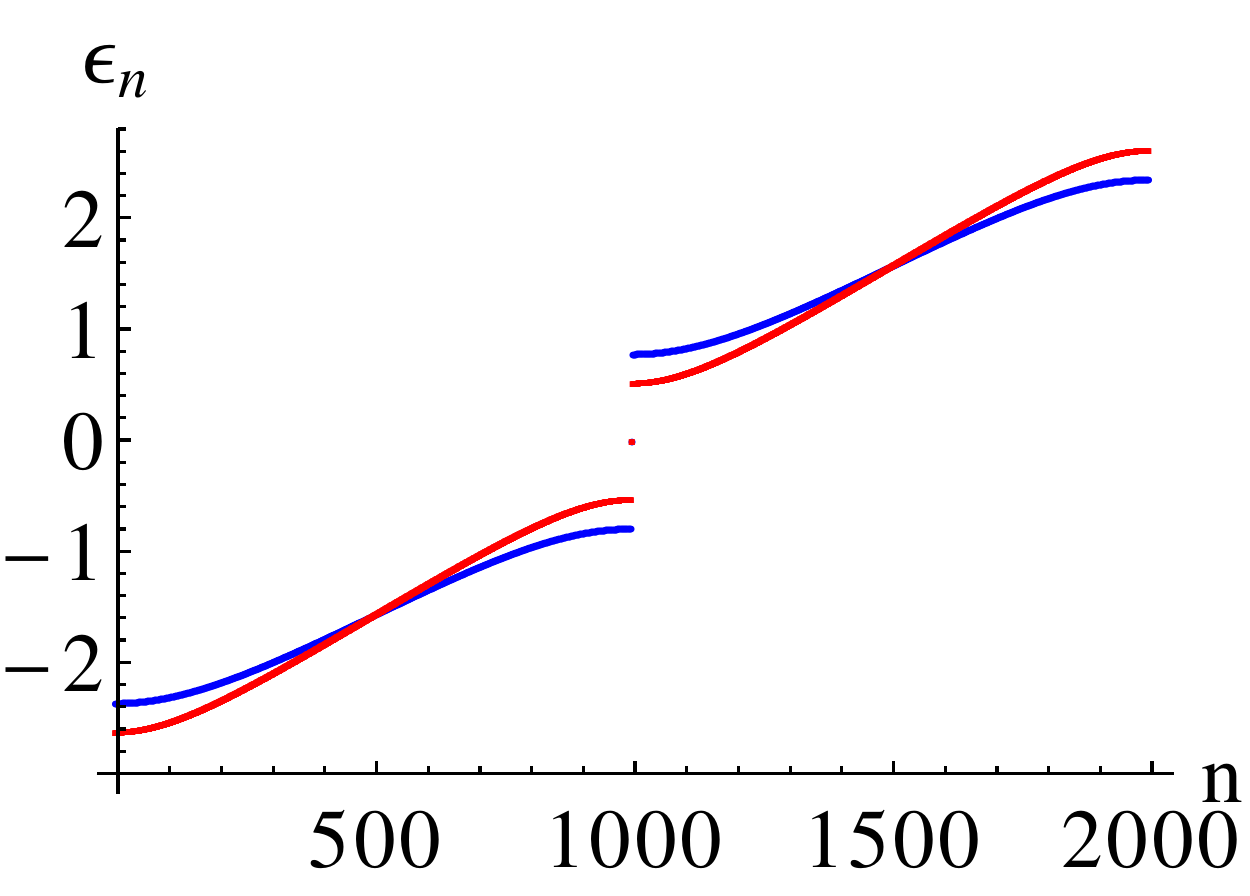}
\caption{(Color online). Spectrum $\epsilon_n=\imath \log(\lambda_n)$
where $\lambda_n$ are the eigenvalues of the operator $ \hat W(\theta_1)$ (blue)
and $W(\theta_2)$ (red) for the case $\theta_1=\pi/4$ and $\theta_2=\pi/6$.}
\label{fig:twospect_time}
\end{figure}

In the case of DTQW with time dependent coin operations given by aperiodic sequences we deal with a 
non-autonomous quantum system and therefore we cannot, strictly speaking,
define a general energy spectrum. However, we can look at  the instantaneous  or the asymptotic energy spectrum. Applying different coin operations at different time steps
amounts to changing the energy spectrum between two possible spectra
in an aperiodic way and the two instantaneous energy spectra
for the cases $\theta_1=\pi/2$ and $\theta_2=\pi/6$
are shown in Fig.~\ref{fig:twospect_time}.

\begin{figure}[tb]
\includegraphics[width=8cm]{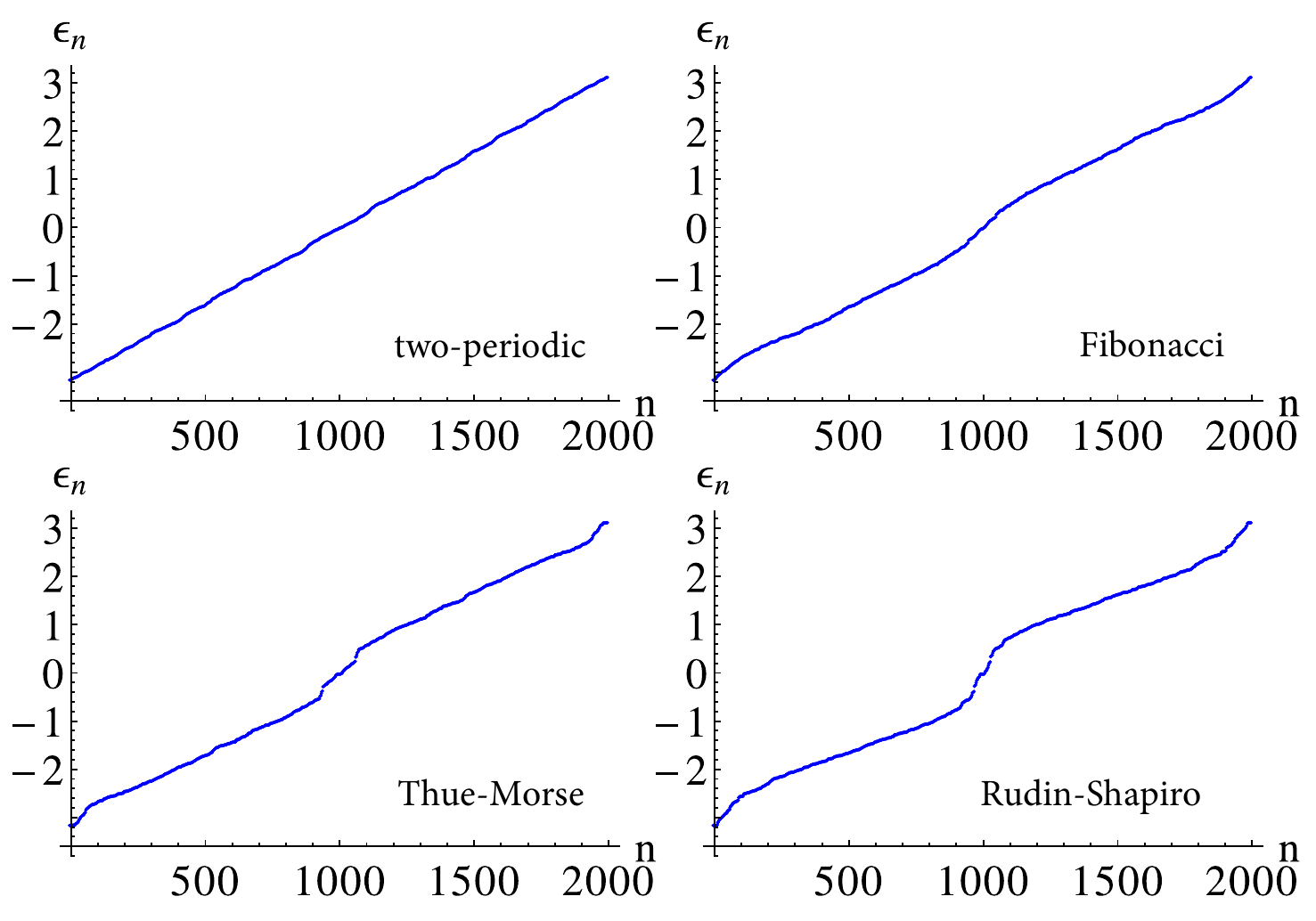}
\caption{(Color online). Lowest band of the spectrum $\epsilon_n=\imath \log(\lambda_n)$
where $\lambda_n$ are the eigenvalues of the operator $ \hat U(t)$
with $t=500$. The angles chosen are $\theta_1=\pi/4$ and $\theta_2=\pi/6$. }
\label{fig:spect_time}
\end{figure}

\begin{figure}[h]
\includegraphics[width=8cm]{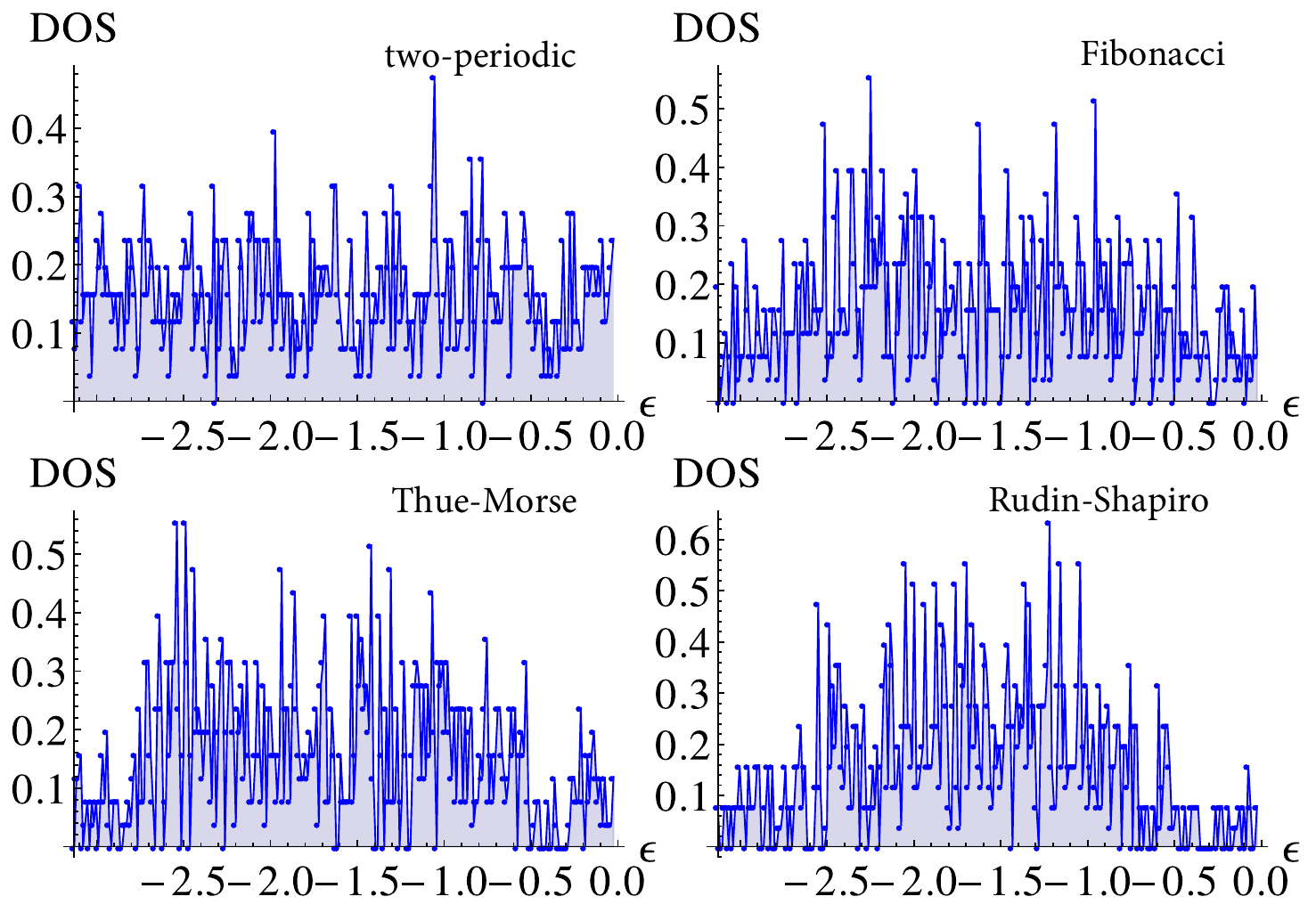}
\caption{(Color online). Densities of states normalized to the total number of states and 
corresponding to the spectra shown in Fig.~\ref{fig:spect_time}. The energy window used to compute it
is the same as the one that will be used for the analysis of the asymptotic properties $d\epsilon=2\pi/t_f$
where $t_f=500$ is the number of steps.}
\label{fig:dos_time}
\end{figure}

To determine the asymptotic properties of the system
it is necessary to look at the spectrum of the total evolution operator $\hat U(t)$ for $t\gg 1$.
We do this numerically and note that after only a few $(t \approx 30)$ steps the spectrum of the total evolution operator does not appreciably change any more. 
%\tr{This means that after $t\approx 30$ steps the system evolves with the same spectrum which is responsible for the asymptotic dynamics of the system.} 
The results for the different sequences are shown in Fig.~\ref{fig:spect_time} and one can  immediately notice  the absence of  sub-band gaps, which characterized the spectra in the spatially dependent case.
On the other hand, the DOS for the spectrum of the asymptotic evolution operator reveals a much richer structure (see Fig.~\ref{fig:dos_time}) than what we could have gathered from the spectrum itself.
In particular one can note features of the self-similarity for the Fibonacci (top-right) and Thue-Morse  (bottom-left) cases. This means that although the single step is completely homogeneous, the aperiodic nature of the temporal distribution arises clearly in the long time limit.

%=====================================
\section{Spreading of the walker}
\label{sec5}
%=====================================

In order to characterize the dynamics of the walker we look 
at both, the mean displacement $\langle x(t)\rangle$ and its variance 
given by $\sigma^2(t)=\langle x^2(t)\rangle-\langle x(t)\rangle^2$.
We choose the initial state to be $\ket{\psi} = \ket{x=0}\otimes\frac{\ket{\uparrow} + i\ket{\downarrow}}{\sqrt{2}}$,
such that the spatial part is localized
and the initial coin state as $\theta_1=\theta_2=\pi/4$. This leads to a balanced DTQW. In the following we will discuss the different sequences in detail.

\begin{figure}[tb]
  \includegraphics[width=\linewidth]{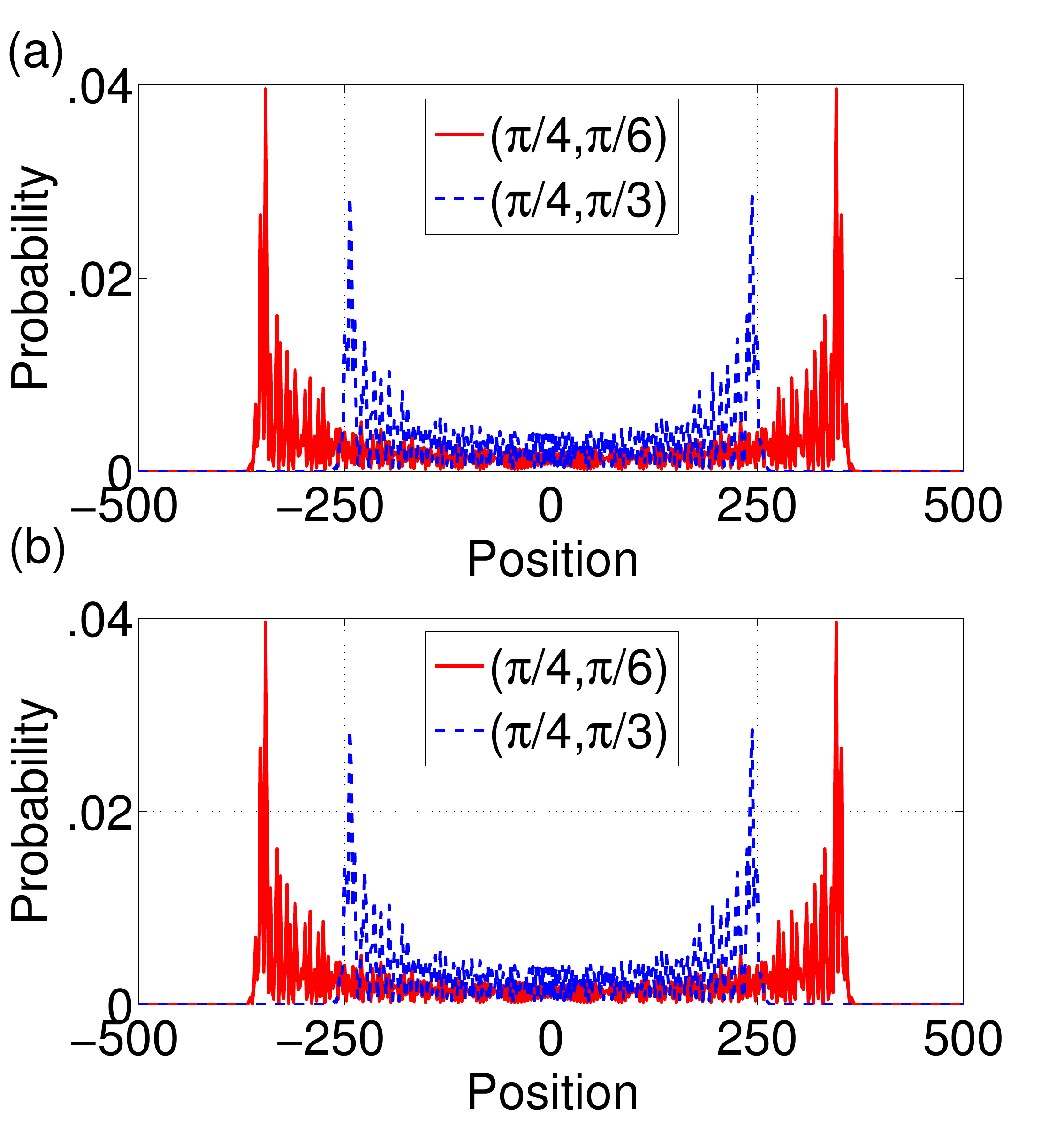}
  \caption{(Color online). Probability distribution after  500 steps of DTQW using a two-periodic coin for two sets of angles $(\theta_1,\theta_2)$. (a) spatial coin dependence (b)  temporal coin dependence. }
  \label{fig:TwoPeriodicSpreading}
\end{figure}

{\it Two-periodic:}
The dynamics of a standard DTQW with a single, fixed coin operator is known to be always ballistic, with a velocity that depends on $\theta$. 
From Fig.~\ref{fig:TwoPeriodicSpreading} one can see that this property prevails for the two-period sequence in the temporal as well as in the spatial case, but
with the velocity $v$  depending now on $\theta_1$ and $\theta_2$. This result is not very surprising, since the presence of a periodicity means that localization is absent and therefore the spread in position space has to have the form that has the characteristic peaks in the outer regions. 
\par 
It is interesting to note that the probability distribution is exactly the same for the spatial and the temporal sequence of the  coin operations (compare Figs.~\ref{fig:TwoPeriodicSpreading}(a) and (b)). This can be understood by a simple symmetry argument: after $t$ steps the possible positions of the walker range from $-t$ to $t$, which means that the total number is always odd. Since the sequence is of period two, all even-numbered positions will have one coin and all odd numbered positions have another coin. The structure of the shift operator ensures that at any given time, the amplitude of the walker is either on the odd positions or on the even positions. Therefore only one coin operation is being applied to the walker at any given step, which establishes the equivalence to the temporal two-period walk.

\begin{figure}[tb]
\includegraphics[width=.96\linewidth]{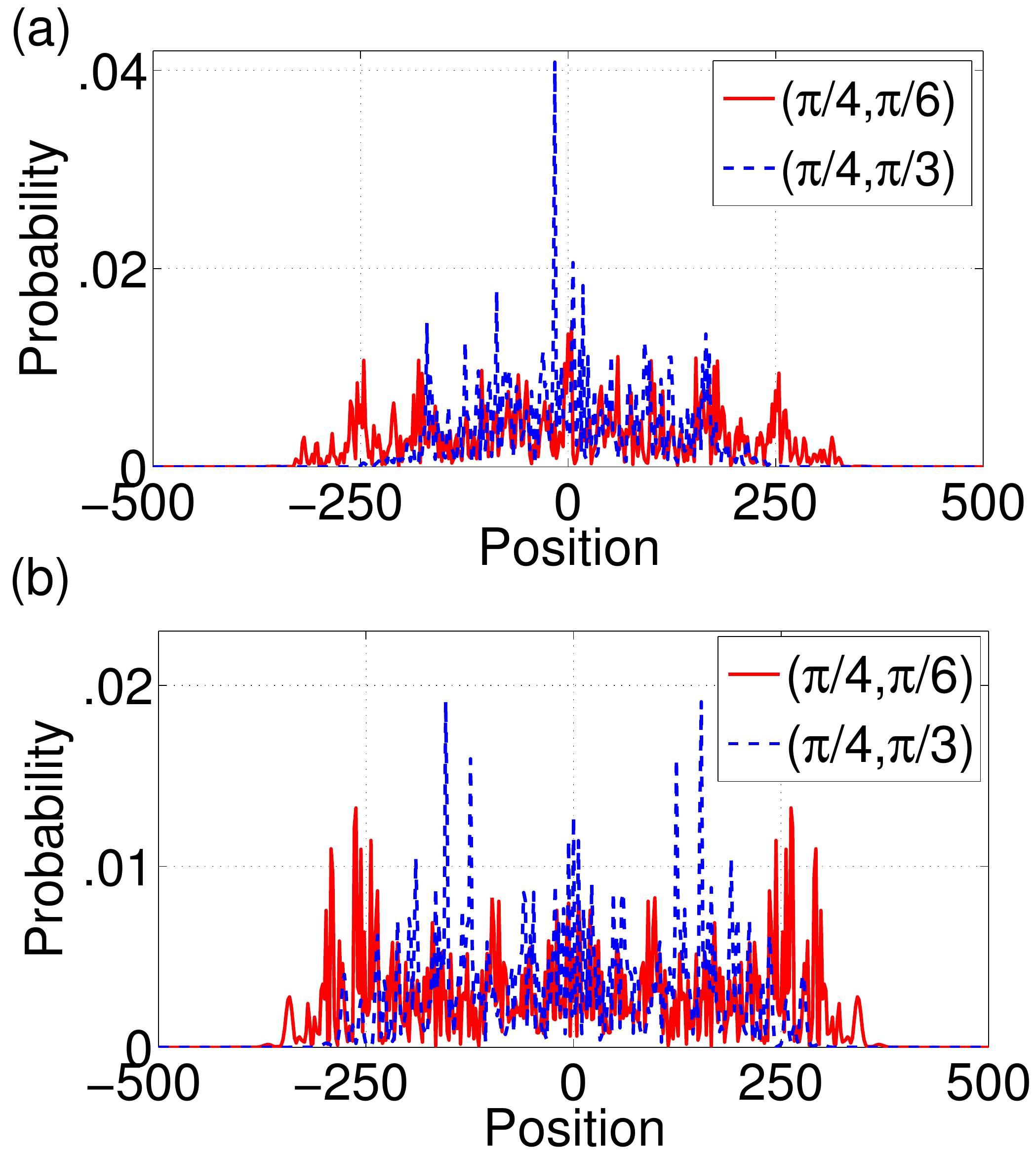}
\caption{(Color online). Same as Fig.~\ref{fig:TwoPeriodicSpreading}, but using a Fibonacci coin. For both situations the distribution can be seen to be dispersive.}
\label{fig:ProbFibonacci}
\end{figure}
{\it Fibonacci:}
The probability distributions in position space after 500 steps with the spatial or temporal arrangement of the coin operation given by the Fibonacci sequence are shown in Fig.~\ref{fig:ProbFibonacci}. It is immediately notable that neither walk features the peaks in the outer regions and that no strongly localized peak  at origin ($x=0$) exists. This indicates that the Fibonacci sequence leads to DTQWs with diffusive behaviour.  

\begin{figure}[tb]
  \includegraphics[width=\linewidth]{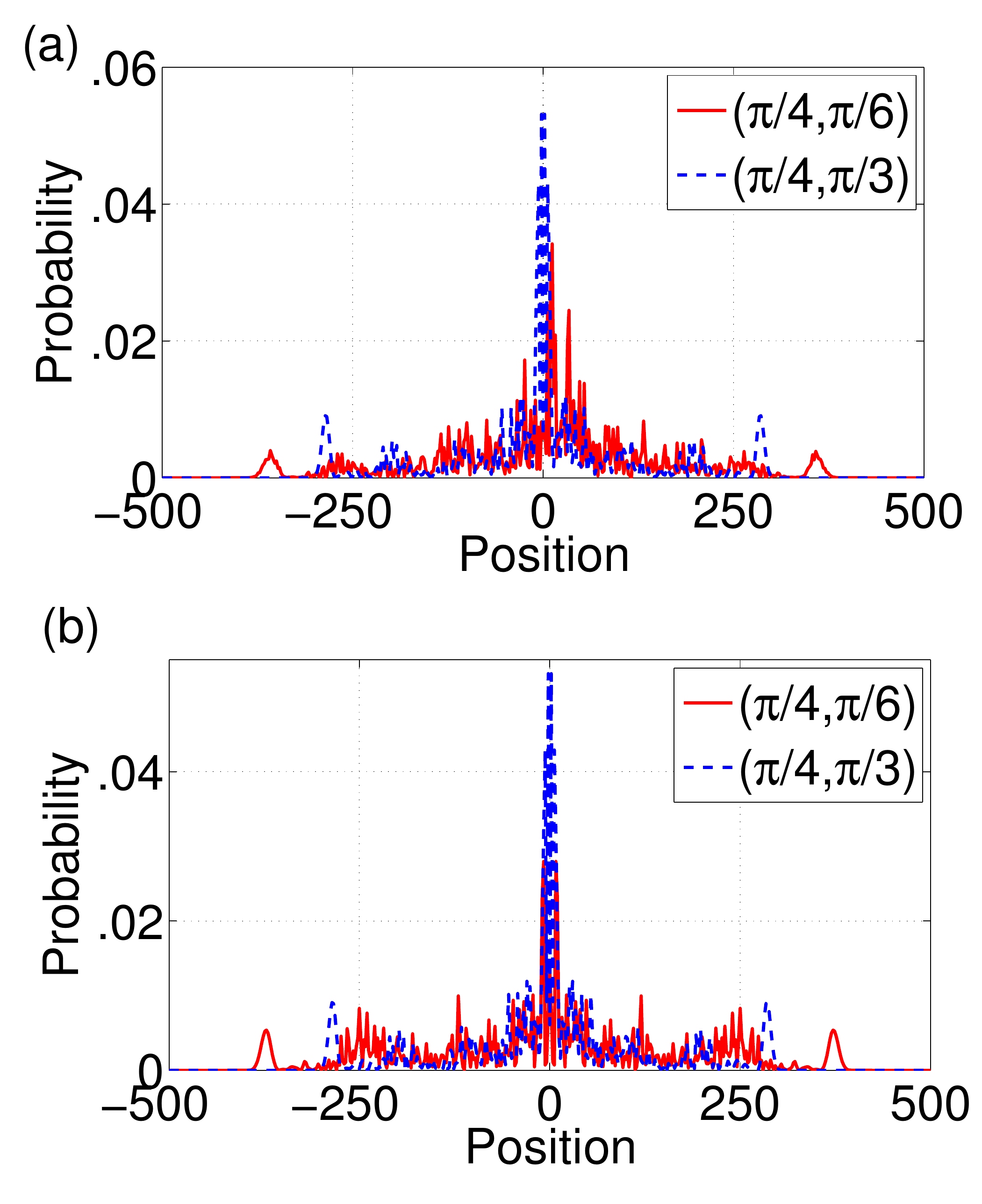}
  \caption{(Color online). Same as Fig.~\ref{fig:TwoPeriodicSpreading}, but using a  Thue-Morse coin. For both situations a localized and a diffusing component are visible in the distribution. }
  \label{fig:ProbTM}
\end{figure}
{\it Thue-Morse:}
The probability distributions for this sequence are shown in Fig.~\ref{fig:ProbTM}. In the spatial and the temporal case a prominent peak localized at $x=0$ can be seen, while at the same time a diffusive component is visible as well.

\begin{figure}[tb]
  \includegraphics[width=.93\linewidth]{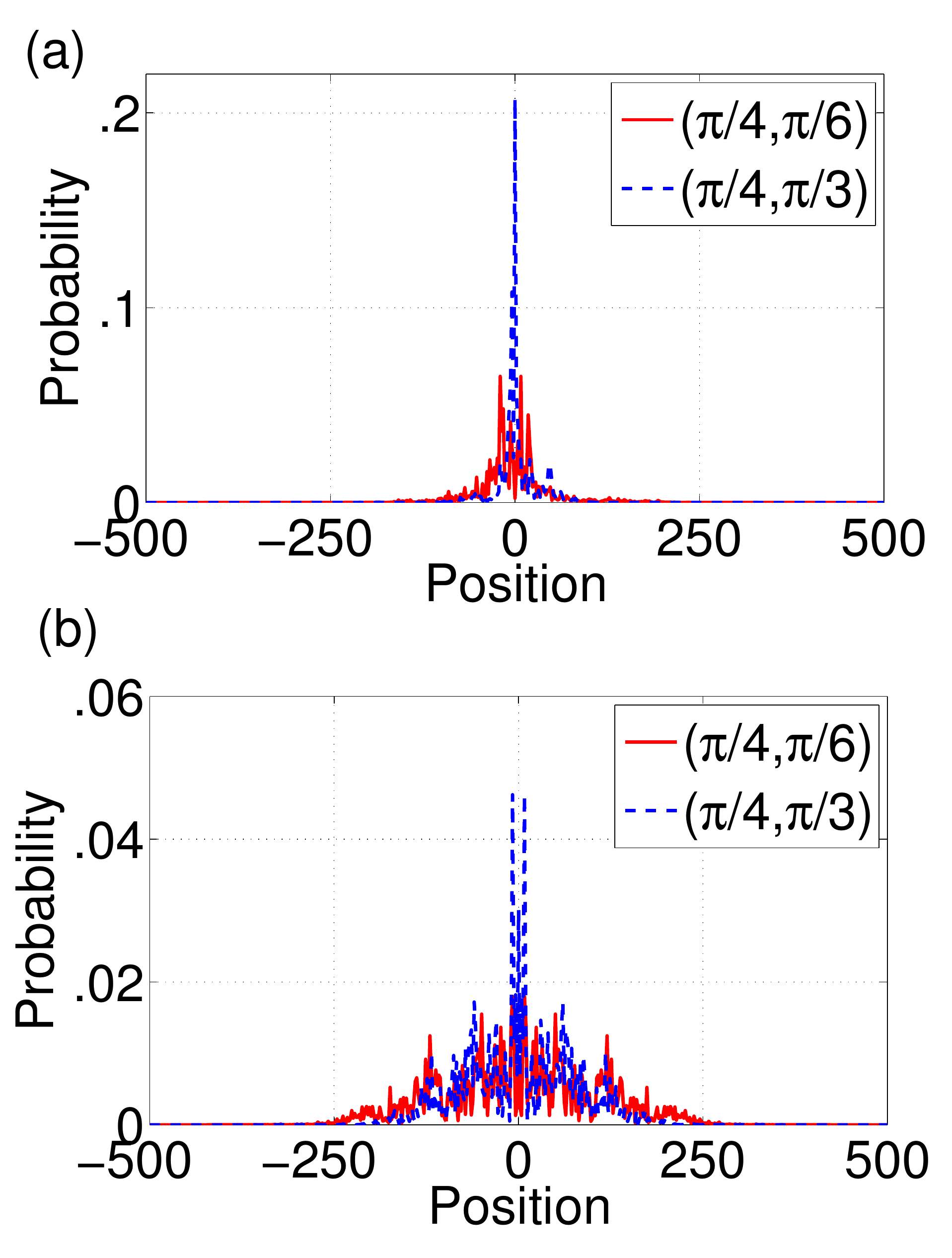}
  \caption{(Color online). Same as Fig.~\ref{fig:TwoPeriodicSpreading}, but using a Rudin-Shapiro coin. The distribution using the spatial sequenence is more strongly localized around the origin than the one for the temporal sequence.}
  \label{fig:ProbRudinShapiro}
\end{figure}

{\it Rudin-Shapiro:}
The probability distributions for this sequence are shown in  Fig.~\ref{fig:ProbRudinShapiro}. One can see that, compared to the spread for the Fibonacci and the Thue-Morse sequences, the probability distribution for the Rudin-Shapiro sequence is more localized around the origin, $x=0$. The localization is also stronger for the spatial sequence than for the temporal sequence. 

To better understand the different dynamics displayed by the aperiodic walks, we will in the following examine the dynamics as a function of time and the energy spectra of these walks in more detail.

%=====================================
\subsection{Position dependent coin operations}
\begin{figure}[h]
  \includegraphics[width=9cm]{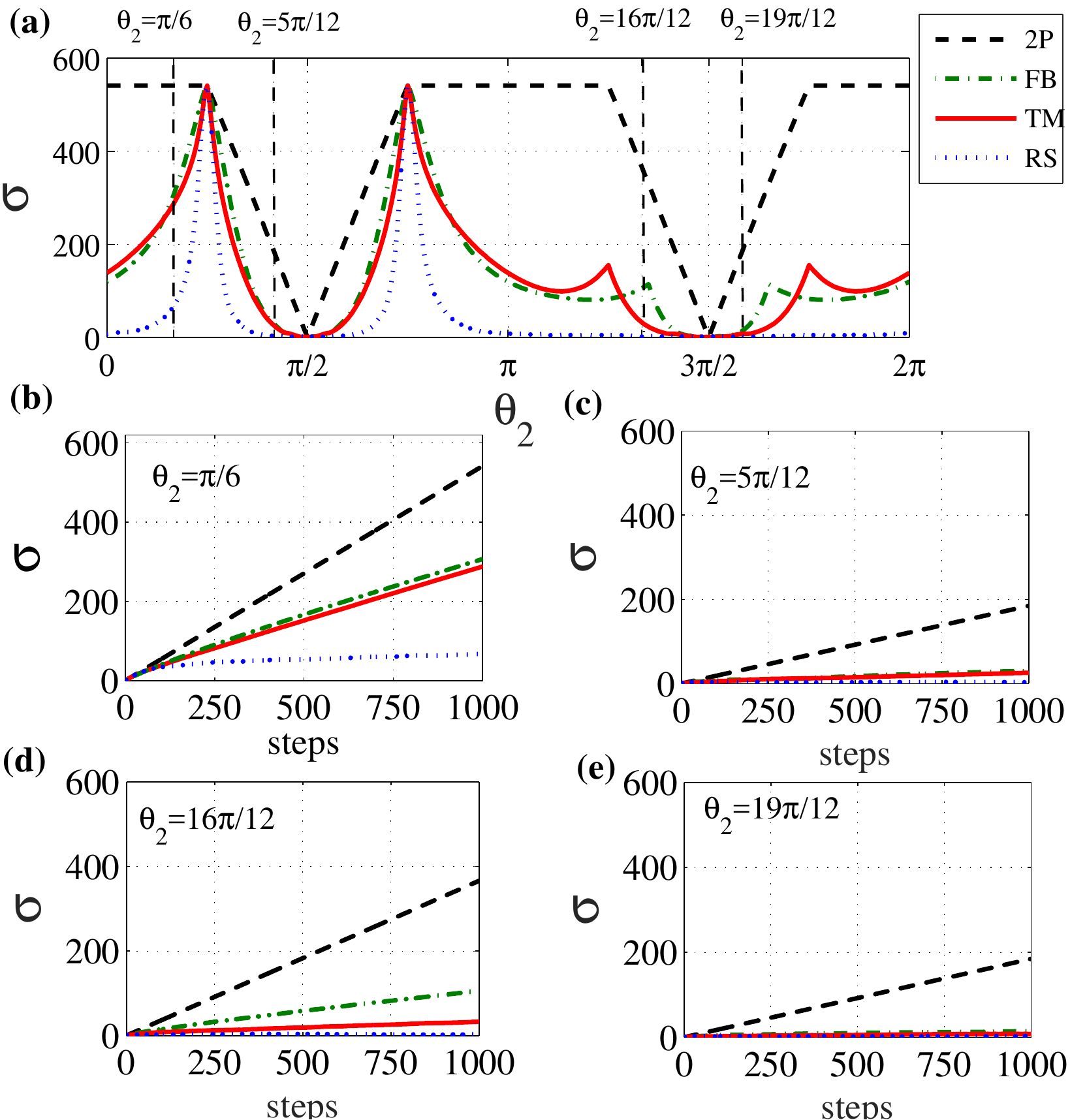}
  \caption{(Color online). Spatially dependent coin. (a) Standard deviation at the end of a 1000 step walk as a function $\theta_2$ with $\theta_1=\frac{\pi}{4}$. (b)-(e) Standard deviation as a function of time for 1000 steps for different values of $\theta_2$. The notation used for the sequences are 2P $\leftrightarrow$ two-periodic,  Fb $\leftrightarrow$ Fibonacci, TM $\leftrightarrow$ Thue Morse, and RS $\leftrightarrow$ Rudin-Shapiro.}
  \label{fig:IDTQW}
\end{figure}

%=====================================
We will first focus on the walks with spatially varying coins and note that while in the standard DTQW evolution the spread of the probability distribution depends on the coin parameter $\theta$~\cite{NV01, CSL08}, its general form is independent of the specific value.  We can therefore restrict the parameter space by choosing one of the coins to be a Hadamard ($\theta_1=\pi/4$) and the results we obtain will hold for all values of $\theta_1$.  
 For all four forms of walks, the standard deviation after 1000 steps and as a function of $\theta_2$ is shown in Fig.~\ref{fig:IDTQW}(a). One can see that the standard deviation for all four sequence is zero when $\theta_2 = \pi/2$ and $3\pi/2$ and maximum and equal  when $\theta_2 = \pi/4$ and $3\pi/4$. For all other values, the standard deviation is always highest for two-period walk and lowest for Rudin-Shapiro sequence .  For the Fibonacci and the Thue-Morse sequences, the standard deviations are in between these two extremes. For a detailed look into the distributions, we have picked four different values of $\theta_2$ and shown the standard deviation as a function of the number of steps in Figs.~\ref{fig:IDTQW}(b)-(e).
 
Let us first look at the Fibonacci sequence, which has a pure-point spectrum, and is therefore the closest to a periodic structure.  Comparing Figs.~\ref{fig:TwoPeriodicSpreading}(a) and ~\ref{fig:ProbFibonacci}(a) one can see that the width of the spread in position space is roughly identical for both.  However, the standard deviations shown in Figs.~\ref{fig:IDTQW}(b)-(e) show a slower increase for Fibonacci sequence compared to two-periodic sequence. In fact, the difference between them grows with increasing number of steps.  Unlike the two-periodic walk, where the peaks in the probability distribution are observed at the outer edges of the distribution, the distribution from the Fibonaacci sequence shows larger peaks at multiple spatial positions. 

For the walk using the Thue-Morse sequence, which has a singular-continuous spectrum, the  probability distribution in Fig.~\ref{fig:ProbTM} shows that the distribution has both a localized and a diffusing components. The standard deviation as function of number of steps shown in Figs.~\ref{fig:IDTQW}(b)-(e) show that it is smaller than the two-periodic sequence and the Fibonacci sequence.  

Finally, the Rudin-Shapiro sequence has an absolutely continuous diffraction spectrum, similar to a completely disordered medium. Here by disorder we mean a random distribution of two values assigned to each point on the lattice.  From Figs.~\ref{fig:ProbRudinShapiro} and \ref{fig:IDTQW}(b)-(e) one can see that the distribution  localizes  around the origin and the value of the standard deviation remains small with increasing number of steps. 

%=====================================
\subsection{Time dependent coin operations}
%=====================================

\begin{figure}[tb]
  \includegraphics[width=9.0cm]{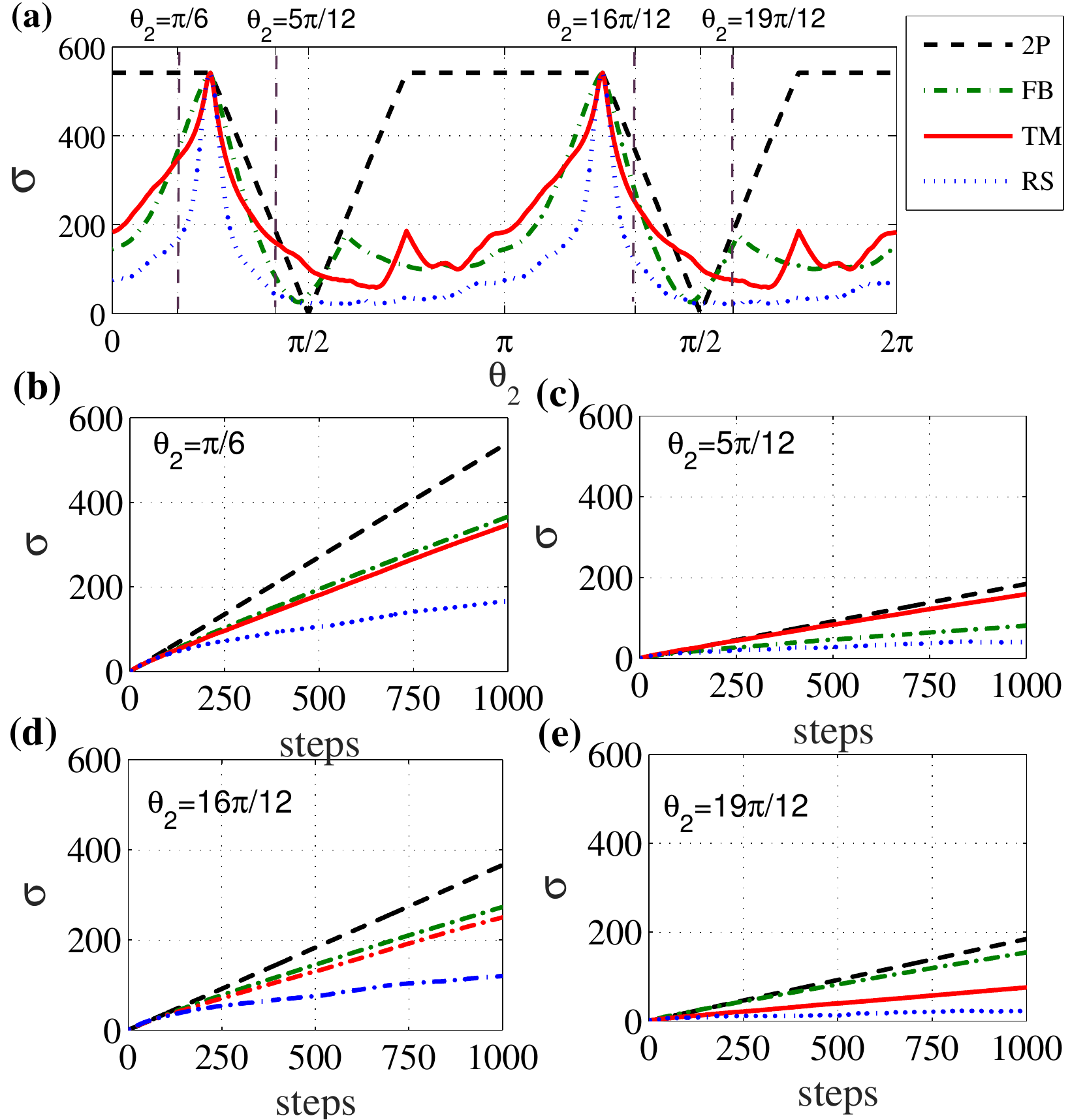}
  \caption{(Color online). Time dependent coin. (a) Standard deviation at the end of a 1000 step walk as a function of $\theta_2$ with $\theta_1=\frac{\pi}{4}$. (b)-(e) Standard deviation as a function of time for 1000 steps for different values of $\theta_2$. The notation used for sequences are the same as in Fig.~\ref{fig:IDTQW}.}
  \label{fig8}
\end{figure}

The standard deviation of a DTQW with time dependent coin operations after 1000 steps and as a function of $\theta_2$ is shown in Fig.~\ref{fig8}(a). One can see that the standard deviation is always lowest for the Rudin-Shapiro sequence, which has the absolutely continuous diffraction spectrum, similar to a completely disordered medium. A detailed look into the development of the deviations is again given for four different values of $\theta_2$ as a function of the number of steps in Figs.~\ref{fig8}(b)-(e).
One can note that the behaviour is qualitatively similar to the case of the spatially dependent coin, with values for the standard deviation generally higher. This can be understood by noting that it takes a larger number of steps for the system to feel the effects of the aperiodicity. 

%=====================================
\section{Asymptotic behavior}
\label{sec:asymptotic}
%=====================================

We now turn to the characterization of the asymptotic properties of the DTQWs
with different coin sequences and  will highlight  
the relation between the asymptotic behavior of the walker distribution
and the spectral properties of the chosen aperiodic sequence.
In order to make this discussion quantitative we will  choose the so-called survival probability or Loschmidt echo $\mathcal{L}(t)=|\nu(t)|^2$  as our figure of merit, which is defined via its amplitude as
\begin{equation}
 \nu(t)=\braket{\psi(0)|\psi(t)},
\end{equation}
where $\ket{\psi(t)}=U(t)\ket{\psi(0)}$.

The reason to consider this quantity is two-fold.
On one hand it has a clear physical meaning because it describes  the probability of finding the evolved state within the initial state
after some time. Therefore, it is strictly linked to the study of ergodicity 
and localization of a quantum system and as such has been widely studied in statistical mechanics.
On the other hand it is directly related to the Fourier transform of the spectral measure of the evolution
operator~\cite{Last1996} (and therefore, for an autonomous system, 
of its infinitesimal generator: the Hamiltonian)
\begin{equation}
|\nu(t)|^2=\left|\int_{\sigma} d\mu_0(\epsilon) e^{-\imath \epsilon t} \right|^2,
\label{eq:ft_nu}
\end{equation}
where $\mu_0$ is the measure induced by the initial state $\ket{\psi(0)}$.
Therefore its Fourier transform is the measure itself.

A second quantity we will explore is the Ces\'aro average (time average) of the echo, which is
defined as
\begin{equation}
\langle|\nu|^2\rangle_T=\frac{1}{T}\sum_{t=0}^{T-1}|\nu(t)|^2.
\label{eq:ca_nu2}
\end{equation}

%=====================================
\subsection{Position dependent coin operations}
%=====================================

\begin{figure}[tb]
\includegraphics[width=8cm]{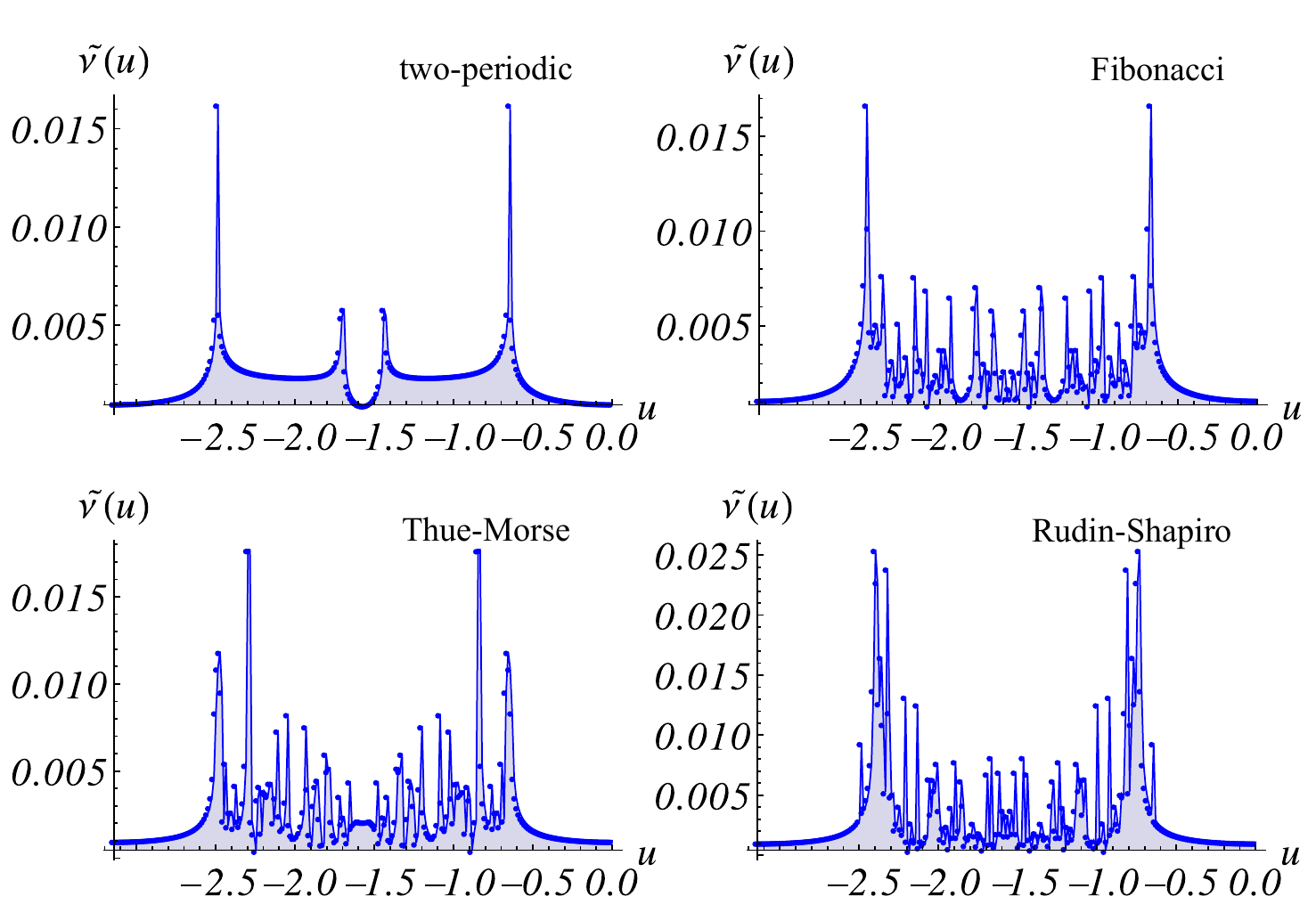}
\caption{(Color online). Fourier transform $\tilde \nu(u)$ of the 
survival probability $\nu(t)$ for DTQWs with spatially inhomogeneous coins.
The angles chosen are $\theta_1=\pi/4$ and $\theta_2=\pi/6$.}
\label{fig:ft_space}
\end{figure}

In the case of position dependent coin operations
the corresponding DTQW is autonomous and therefore
its asymptotic properties are related to the system's energy spectrum (and to the initial state).
We will consider an initial state such that the walker is localized at the
centre of the lattice $\ket{x=0}$ and the coin state is $\ket{\uparrow}$.
Since this initial state is localized in real space it is completely delocalized
in momentum space and therefore we expect it to have a non-vanishing overlap
with nearly all eigenstates of the evolution operator.
For the same reason we have choosen an asymmetric coin state.
It therefore allows us to explore the full spectrum of the system.
Indeed, the peaks in the Fourier transforms $|\tilde \nu(u)| $ of the survival amplitude
in Fig.~\ref{fig:ft_space} closely resemble those of the DOS shown in Fig.~\ref{fig:dos_space}.
This confirms that the chosen initial state is such that 
each eigenstate is initially occupied to some extent
and we can therefore explore the full energy spectrum during the dynamics.

Looking closer at 
 the Fourier transforms shown in Fig.~\ref{fig:ft_space} one can note in all of them
 sharp peaks around $u\approx -2.5$
and $u\approx -0.7$. These peaks are due to the sub-bands generated
by the presence of the internal degree of freedom (coin), which
can be confirmed by comparing with the corresponding spectra shown in Fig.~\ref{fig:inhom_spect}.
This is also confirmed by looking at the plot for the two-periodic case
where two additional sharp peaks appear at $u\approx -1.7$ and $u\approx-1.4$
which correspond to the further breaking of the corresponding spectrum
into new sub-bands due to the modulation of the operators.
On the other hand, as expected, the remaining part is continuous and has no sharp peaks.

Looking at $\tilde \nu(u)$ for the aperiodic arrangements
we can see that, beside the above mentioned sharp peaks corresponding
the breaking of the spectrum into two main sub-bands,
a series of sharp peaks exist whose positions are aligned with those of the 
corresponding DOS shown in Fig.\,\ref{fig:dos_space}.
The supports of these peaks belongs to the discrete part of the spectrum
\cite{Last1996} and we will in the following determines  whether they are
pure point or singular continuous.

Given that  the survival amplitude in Eq.\,(\ref{eq:ft_nu}) is the Fourier transform of the measure, one can identify
 a singular continuous spectrum whenever the conditions 
(compare with Eq.(2) in Ref.\,\cite{Pikovsky1995} for the averaged 
correlation function of an observable $f(t)$)
\begin{subequations}
\begin{align}
&\lim_{t\rightarrow \infty}\nu(t)\neq 0 \hspace{0.5cm} &&\text{not absolute continuous}
\label{eq:RAGE}\\
&\lim_{T\rightarrow \infty}\langle|\nu|^2\rangle_T=0 && \text{not pure point}
\label{eq:wiener}
\end{align}
\end{subequations}
are both satisfied. The first condition ensures the absence of an absolutely continuous part
in the spectrum because, according to the RAGE theorem~\cite{Last1996},
$\lim_{t\rightarrow \infty}\nu(t)= 0$ is  a necessary condition for the presence of an absolutely continuous
part in the spectrum. The second condition ensures the absence of pure point spectrum
due to Wiener's lemma (see lemma 1.1 in \cite{Queffelec2010}).
Therefore the simultaneous occurrence of conditions in Eqs.\,(\ref{eq:RAGE}) and (\ref{eq:wiener})
can be interpreted as the spectrum having only singular continuous parts.

\begin{figure}[tb]
 \includegraphics[width=\linewidth]{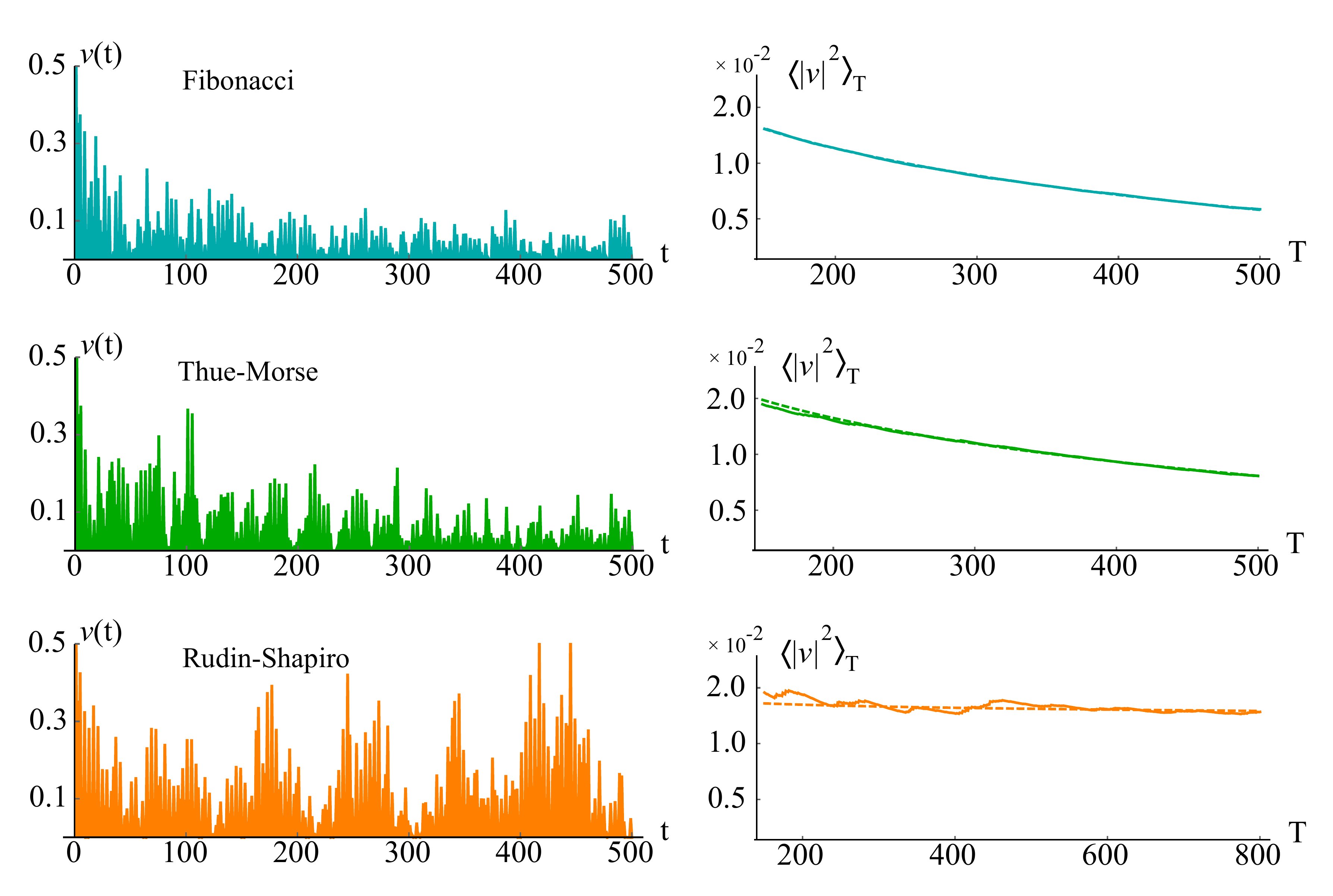}\\
\caption{(Color online). Left: Survival amplitude $\nu(t)$
for different distribution of the coin operators in a DTQWs with spatially aperiodically distributed coin operations.
Right: Corresponding Ces\'aro averages of $|\nu(t)|^2$.
The dashed lines are the fit functions.}
\label{fig:nu_space}
\end{figure}

On the other hand, the above discussion has shown that in all cases at least four delta-like peaks, corresponding to the splitting of the band into two sub-bands, can be expected and therefore the spectrum has a non-empty discrete component. Bearing this in mind we show in Fig.~\ref{fig:nu_space} the asymptotic values of  $\nu(t)$  and 
the corresponding Ces\'aro averages,  $\langle|\nu|^2\rangle_T$ . 

It is clear that $\nu(t)$ goes to zero as $t\rightarrow \infty$ only for the two-periodic case.
This can be interpreted as the spectrum having an absolutely continuous part (RAGE theorem),
which correspond to the continuum inside each of the four sub-bands.
In all other cases $\nu(t)$ does not asymptotically vanish  and one can therefore expect that the corresponding spectra lack 
an absolutely continuous part. We notice that among all the Rudin-Shapiro sequence is the one which
maintains the highest value at infinite times. The  Ces\'aro averages for all sequences do not go to zero over the simulated time, which 
suggests the presence of discrete parts in the specta and which are due to the splitting into subbands.
In order to estabilish whether or not the Ces\'aro averages tend to zero at infinity we have fit the data with a power-law fuction $f(T)=T^\alpha$ with $\alpha<0$ for the Fibonacci and Thue-Morse cases,
whereas for the Rudin-Shapiro we found 
that the best agreement is with a fitting function of the form $f(T)=e^{\alpha t^{\beta}}$
with $\alpha<0$ and $0<\beta<1$.
Note that, in order to counter the localization effects from the Rudin-Shapiro 
sequences, we have simulated a larger number of steps to get a better accuracy
for the parameters $\alpha$ and $\beta$.
As it can be seen from the right hand side of Fig.\ref{fig:nu_space} in the Fibonacci and 
Thue-Morse cases the fit gives a very good agreement and the Ces\'aro average 
which decays algebraically to zero. On the other hand for the Rudin-Shapiro case  the 
decay is slower due to localisation effects and in particular we found $\beta=(0.013\pm 0.001)$.
This localisation might be useful for applications such as
quantum memories using DTQW \cite{CB15}.

Due to the asymptotic decay of the Ces\'aro average
we can therefore conclude that the spectrum 
of a DTQW with spatially distributed coin operations 
has a singular continuous component only (beside the 
discrete component introduced by the presence of the internal degree of freedom
of the coin).

\begin{figure}[tb]
\includegraphics[width=\linewidth]{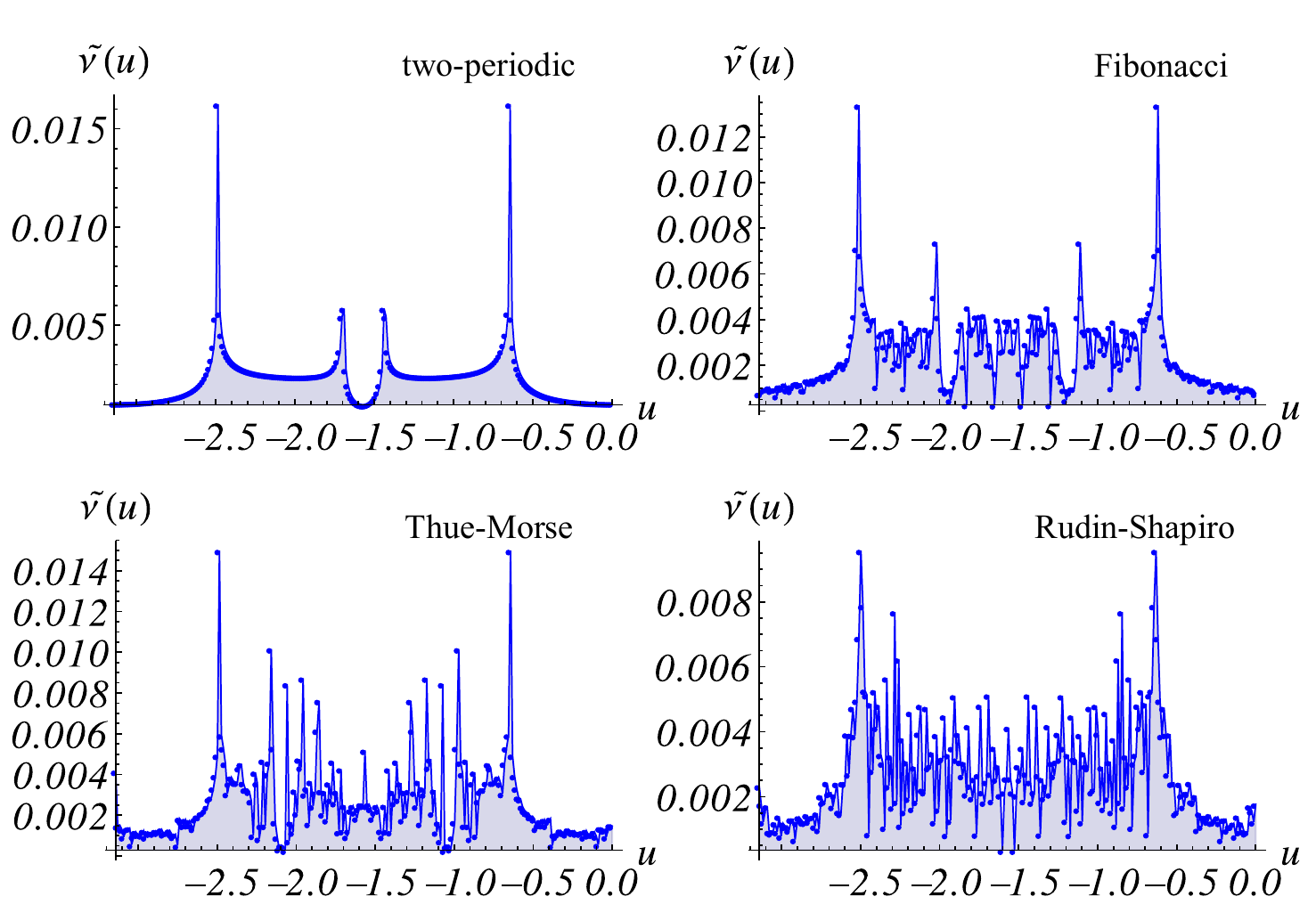}
\caption{(Color online). Fourier transform $\tilde \nu(u)$ of the 
survival probability $\nu(t)$ for temporally inhomogeneous DTQWs.
The angle chosen are $\theta_1=\pi/4$ and $\theta_2=\pi/6$.}
\label{fig:ft_time}
\end{figure}

%============================================
\subsection{Time dependent coin operations}
%============================================
In the case of time dependent coin operations
the asymptotic behavior is characterized 
by the spectrum of the asymptotic evolution operator
as discussed in Sec.~\ref{sec:en_spect}.
The Fourier transform of the survival amplitude $\tilde \nu(u)$
is shown in Fig.\,\ref{fig:ft_time} and has a  form similar to the spatial case discussed above. 
However,
the peaks in $\tilde \nu(u)$
are not similar to the ones in the corresponding DOS of the asymptotic
evolution operator, which is shown in Fig.~\ref{fig:dos_time}.
Nevertheless, signatures of the presence of aperiodic order is clearly visible in the 
sharp  spikes and in the self-similar character of $\tilde \nu(u)$.

\begin{figure}[tb]
\begin{tabular}{cc}
 \includegraphics[width=\linewidth]{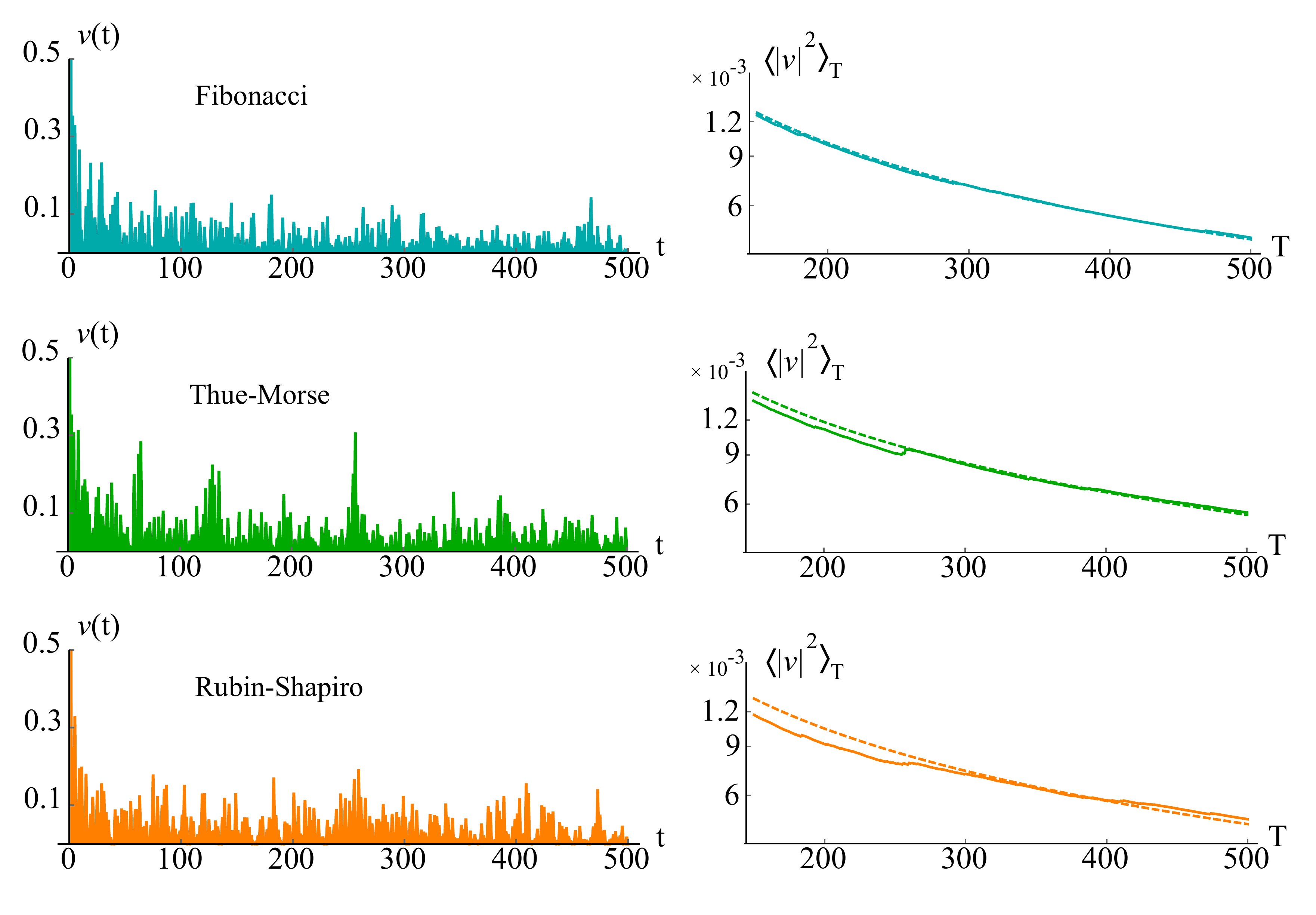}\\
\end{tabular}
\caption{(Color online). Survival amplitude $\nu(t)$
for different distribution of the coin operators in a DTQWs with spatially aperiodically distributed coin operations.
Right: Corresponding Ces\'aro averages of $|\nu(t)|^2$.
The dashed lines are the fit functions.}
\label{fig:nu_time}
\end{figure}

It is therefore interesting to again look at the behavior of
the survival amplitude and its Ces\'aro average.
From the left hand side of Fig.~\ref{fig:nu_time} we see that 
the survival amplitude does not go to zero for the large times
considered and for all sequences, which allows us
to conclude that the corresponding energy spectra
do not possess an absolutely continuos part.

Looking at the Ces\'aro average (right hand side of Fig.~\ref{fig:nu_time}) we observed that in the temporal case 
the best fit is a power-law of the form $f(T)=\alpha T^\beta$ for all instances. 
It is interesting that the Rudin-Shapiro case, which 
was the slowest to converge to zero for large $T$ in the spatial case,
now behaves similarly to the more delocalized walks.
Indeed, as it can be observed in Fig.\,\ref{fig:ProbRudinShapiro},
for the case of the  temporal Rudin-Shapiro distribution of the coin operations, the 
wave-packet spreads more than in the case of a spatial Rudin-Shapiro distribution.
Therefore, one can conclude that also for temporal distributions the system 
has a singular continuos spectrum.

%==========================
\section{Conclusions}
\label{conc}
%==========================

In this work we have presented the study of DTQW with different configuration of aperiodic sequences of position and time dependent coin operations. Analysing  the dynamics of the different  aperiodic walks by studying the energy spectrum, the standard deviation of the probability distribution, and their asymptotic behavior we have shown that the DTQW using a Fibonacci sequence, which has pure point spectrum, and a Rudin-Shapiro sequence, which has continuous spectrum, are closer to DTQW with periodic sequence (diffusive spreading) and random sequence (localization), respectively.  Quantum walks using Thue-Morse sequences, which have a singular-continuous spectrum, comprise of both, diffusing and localized component in the dynamics. This establishes that quantum walks are an interesting playground to study the interplay between singular continuous spectra and quantum effects and in particular their competition in the transport properties in quantum systems.

\section{Acknowledgments}
NLG thanks F. Plastina for useful discussions.
NLG and LD acknowledge financial support by MIUR through FIRB 
Project No. RBFR12NLNA002.
NLG acknowledges financial support by the EU Collaborative project 
QuProCS (Grant Agreement 641277).
CMC acknowledge support from DST through Ramanujan Fellowship grant nunber SB/S2/RJN-192/2014. This work was supported by the Okinawa Institute of Science and Technology Graduate University.

%\clearpage

\end{document}